\documentclass{article}
\usepackage[square,sort,comma,numbers]{natbib}
\usepackage{lipsum}
\usepackage{xcolor}

\newcommand{\blue}[1] {{\color{blue} #1}}

\usepackage{graphicx}
\usepackage{amsmath}
\usepackage{amssymb}

\usepackage{verbatim}
\usepackage{wasysym}
\usepackage{url}
\graphicspath{{Pics/}}
\usepackage{booktabs}
\usepackage{makecell} 
\usepackage{hyperref}
\usepackage{ragged2e}
\usepackage[a4paper,top=2cm,bottom=2cm,left=2cm,right=2cm,marginparwidth=1.75cm]{geometry}

\begin{document}

\title{High Strain Rate Compressive Deformation Behavior of Nickel Microparticles}

\maketitle

\author{Bárbara Bellón$^{1*}$}, \author{Lalith Kumar Bhaskar$^{1}$}, \author{Tobias Brink$^{1}$}, \author{Raquel Aymerich-Armengol$^{1,2}$}, \author{Dipali Sonawane$^{1}$}, \author{Dominique Chatain$^{3}$}, \author{Gerhard Dehm$^{1}$}, \author{Rajaprakash Ramachandramoorthy$^{1*}$}

Corresponding Authors: b.bellon@mpie.de, r.ram@mpie.de\\
1. Max Planck Institute for Sustainable Materials, 40237 Düsseldorf, Germany\\
2. Current address: Center for Visualizing Catalytic Processes, Department of Physics, Technical University of Denmark, 2800 Kgs. Lyngby, Denmark\\
3. Aix-Marseille Univ. CNRS, CINAM, Marseille 13288, France



\justifying

\begin{abstract}

Understanding the mechanical properties of metals at extreme conditions is essential for the advancement of miniaturized technologies. As dimensions decrease, materials will experience higher strain rates at the same applied velocities. Moreover, the interplay effects of strain rates and temperatures are often overlooked and could have critical effects in applications. In this study, for the first time, the rate-dependent and temperature-dependent mechanical response of nickel microparticles have been investigated. The microparticles were obtained by solid-state dewetting of nickel thin films deposited on c-sapphire. They exhibit self-similar shapes with identical sets of planes, facilitating straightforward comparison between particles. This research represents the first in-depth analysis of the mechanical properties of nickel single crystal dewetted microparticles across six orders of magnitude at room temperature and three orders of magnitude at 128 K. Molecular dynamic simulations (MD) were conducted in parallel on particles with the same faceting. In this work, the gap between experiments and simulations has been reduced to over one order of magnitude in size and 3 orders of magnitude in the strain rates. The thermal activation parameter analysis and MD simulations were employed to ascertain whether homogeneous or heterogeneous dislocation nucleation was the dominant mechanism controlling deformation in the particles.
\end{abstract}


\section{Introduction}
Miniaturized testing of materials has seen an increase in the last few decades \cite{Uchic2005_pillars,Oh2009,Jennings2011,Ramachandramoorthy2016,Peng2013,Sharma2018}. It is well known that decreasing sample size promotes changes in deformation mechanisms, typically leading to stronger materials. Moreover, with miniaturization materials will be subjected to higher strain rates in applications with the same impact velocities. These aspects have a significant impact on applications like cold spray, where microstructural changes due to impact speeds and temperature can affect the adhesion and properties of coatings \cite{Reiser2023,Hassani-Gangaraj2018,HASSANI2020198}. Additionally, small-scale particles can enhance corrosion resistance when used as coatings \cite{Jain2020}. Small-scale metals are also being used in medicine and theranostics \cite{Yun2019theranostics,ASLAM2023theranostics} as biochemical sensors \cite{El-Ansary2010} and in flexible sensors \cite{Segev-Bar2013,Costa2019,Liu2023,Luo2023}.

\medskip

Traditionally, compressive micromechanical testing in the literature has used focused ion beam (FIB) based processing techniques to create micropillars \cite{Uchic2005_pillars, Jennings2011, Wehrs2015, Kiener2006} or thin specimens \cite{FIELD200423}. However, this approach is time-consuming and has detrimental effects on the mechanical properties of materials due to Ga$^+$ ion implantation during the milling process and the subsequent creation of dislocation loops \cite{KIENER2007262, Lee2011}. Other approaches to create microscale samples include whisker growth \cite{Brenner1957,Yoshida1968_whiskers}, lithography or selective etching \cite{Soler2012, Bei2007}, nanoimprint \cite{Becker1986, Jacobo-Martin2021} or micro additive manufacturing \cite{Schwiedrzik2022, Ramachandramoorthy2022}. These techniques can produce a large number of self-similar structures. However, lithography techniques typically only allow 2.5D shapes and involve different complex processes. 2.5D refers to structures that have depth but with features that are primarily defined in two dimensions, such as patterns etched into a surface with varying depths, rather than fully three-dimensional freestanding structures. The other additive micromanufacturing methods, though capable of creating 3D structures, are limited to materials that can be etched or electrodeposited. 

\medskip

\begin{figure}[ht]
  \includegraphics[width=0.5\linewidth]{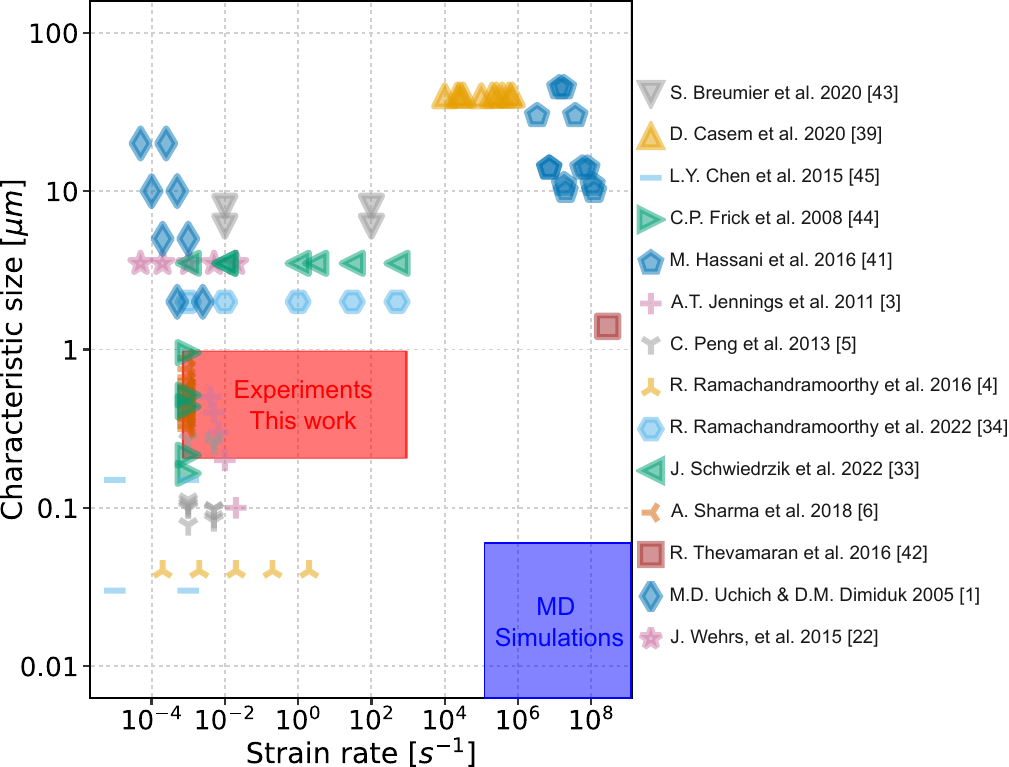}
  \caption{Log-Log plot of the characteristic size in $\mu \mathrm{m}$ vs. strain rates in $\mathrm{s^{-1}}$ of micromechanical tests on FCC metals in literature. Each individual point is a test or an average of tests (depending on the availability of the data in the literature). The blue box represents the bounds of typical molecular dynamics simulations and the red box represents the bounds of the experiments done in the present study.}
  \label{fig:1}
\end{figure}

 Solid-state dewetting (SSD) is an alternative for creating microscale samples. SSD occurs when a continuous thin metal film breaks up into isolated islands or particles when heated below its melting point. This phenomenon is driven by the reduction of surface and interface-free energies. With enough time to equilibrate, it results in small, near-pristine faceted particles \cite{Thompson2012dewetting,Mordehai2011}. Dewetting, usually undesirable for microelectronics due to disconnection of electrical paths, can be leveraged to create nanomaterials or modify surfaces \cite{gentili2012}. Understanding the properties of the dewetted microparticles will allow further developments in this direction. For instance, strains, deformation and dislocations can increase the catalytic activity of nanoparticles, thus making the process more cost-effective \cite{Wang2023catalysis,Portal2021,Chen2020catalysis}. While this process allows the manufacture of numerous particles, minimal changes in pressure, purity of the atmosphere, purity of the film or even contamination in the furnaces influence the surface and interface energies and can thereby dramatically change the shapes and faceting of the particles \cite{Meltzman2011}. Nonetheless, particles obtained by SSD have been tested in literature to study the mechanical properties using conventional micromechanical testers up to quasi-static strain rates of 0.001 $\mathrm{s^{-1}}$ \cite{Sharma2018}. These tests have provided a new insight into how near-pristine materials behave, showing astonishing results in strength and allowing closer comparison with molecular dynamics (MD) simulations. However, the strain rates used on those tests are far away from the limits of MD simulations, highlighted in blue in \blue{\textbf{Figure \ref{fig:1}}}. 
 
 \blue{Figure \ref{fig:1}} shows a representative picture of the micromechanical tests performed in literature on FCC metals with their characteristic sizes and strain rates. It should be noted that this is not an extensive review and the aim is not to gather all the articles published in FCC metal micromechanics but rather to show the gaps in knowledge regarding strain rates. As observed, the majority of the tests are clustered at low strain rates ($<$ 10 $\mathrm{s^{-1}}$) and ranging sizes from tens of nm to tens of $\mu \mathrm{m}$. This is primarily due to the speed limitation of traditional strain gauge load cells. Though they have high load sensitivities, under high-speed testing conditions ($>$ 10 $\mathrm{s^{-1}}$), owing to their low stiffness, resonance frequencies are typically excited and load ringing results \cite{Guillonneau2018}. In order to overcome this problem, different approaches have been reported in literature. For instance, a miniaturized split-Hopkinson pressure bar (SHPB) can be used to compress pillars ($>50~\mu \mathrm{m}$) at strain rates as high as 100,000 $\mathrm{s^{-1}}$\cite{Casem2020}. With this approach, however, constant strain rates during the testing cannot be maintained, nor can the intermediate strain rates in between the limits of the traditional load cells ($\sim$ 0.1 $\mathrm{s^{-1}}$) be accessed. A well-established approach to conduct impact testing that arose over a decade ago is the laser-induced projectile impact testing (LIPIT) \cite{Lee2012}, where microparticles can be shot at surfaces at high velocities. Hassani et at. \cite{hassani2016supersonic} and Thevamaran et al. \cite{Thevamaran2016}, reported the hardness of different FCC metallic particles or nanocubes (respectively) at strain rates $>$ 1,000,000 $\mathrm{s^{-1}}$. The major disadvantage of LIPIT is that extraction of stress-strain signatures is not possible, since only the energies involved in the impact can be calculated based on inbound and outbound velocities. In contrast, a novel system based on piezoelectric load sensors has been developed recently, which by its intrinsically high rigidity and fast response of the piezoelectric materials enables micromechanical testing from quasi-statics to high strain rates up to 1,000 $\mathrm{s^{-1}}$ \cite{Schwiedrzik2022,Ramachandramoorthy2022,Breumier2020}. 

\medskip

From a materials science point of view, lowering the scale of materials typically increases strength due to fewer intrinsic defects and changing deformation mechanisms. The behavior of single crystal FCC metals at small scales and quasi-static speeds has been extensively reported in Refs.~\cite{Uchic2005_pillars,Jennings2011,Peng2013,Frick2008,Chen2015}, among others. There are some studies reporting silver nanocubes compressed at strain rates of nearly $10^8$ $\mathrm{s^{-1}}$, showing nonuniform deformation and martensitic transformation \cite{Thevamaran2016,Thevamaran2020}, that recovers back to its FCC nature via static recrystallization over time. At the bulk scale, polycrystalline FCC metals typically exhibit an uptick in strain rate sensitivity with strain rates beyond 1000 $\mathrm{s^{-1}}$. After a critical strain rate is achieved, a change in the deformation mechanisms occurs where viscous drag forces acting on the dislocations control the deformation rather than typical dislocation gliding \cite{Follansbee1988,Gray1997,MEYERS20031211,Rajaraman2013,Fan2021}. Additionally, in 3D-printed copper micropillars a dependency on the strain rate sensitivity of the grain size beyond a strain rate of 1 $\mathrm{s^{-1}}$ has been identified. This will cause heterogeneous nucleation of dislocations from the surface in the case of big grains ($\sim$ 410 nm) and collective nucleation of dislocations causing saturation in strain rate sensitivity in samples with small grains ($\sim$ 170 nm)\cite{Ramachandramoorthy2022}. Such rate sensitivity transitions are absent in nickel template-assisted electrodeposited micropillars with a grain size of $\sim$ 30 nm \cite{Schwiedrzik2022}. This highlights that information on the deformation behavior of FCC materials at high strain rates is scarce \cite{Follansbee1991} and the effects of the interplay between crystallite size, strain rates, and temperatures are still not well understood. This is what we aim to elucidate in this work using nickel single crystals as a prototypical example.

\medskip

In this work, we report for the first time a thorough investigation of the mechanical properties of nickel single crystal dewetted microparticles across six orders of magnitude in strain rates (0.001–1000~$\mathrm{s^{-1}}$) and two different temperatures: room temperature ($\sim$ 300 K) and cryogenic temperatures (128 K) in combination with finite element method (FEM) based stress-distribution analysis and MD simulations. The microparticles have been obtained by SSD of sputtered nickel thin films. The micromechanical testing has been performed using a custom-modified piezo-based micromechanical testing platform (Alemnis AG) that helped to improve the signal-to-noise ratio at high strain rates. Finally, post-mortem microstructural analysis of the mechanically tested nickel microparticles has been performed in the (scanning) transmission electron microscope ((S)TEM) to explain the observed results.

\medskip

\section{Results and Discussion}
\subsection{Dewetting}

\begin{figure}[ht]
  \includegraphics[width=\linewidth]{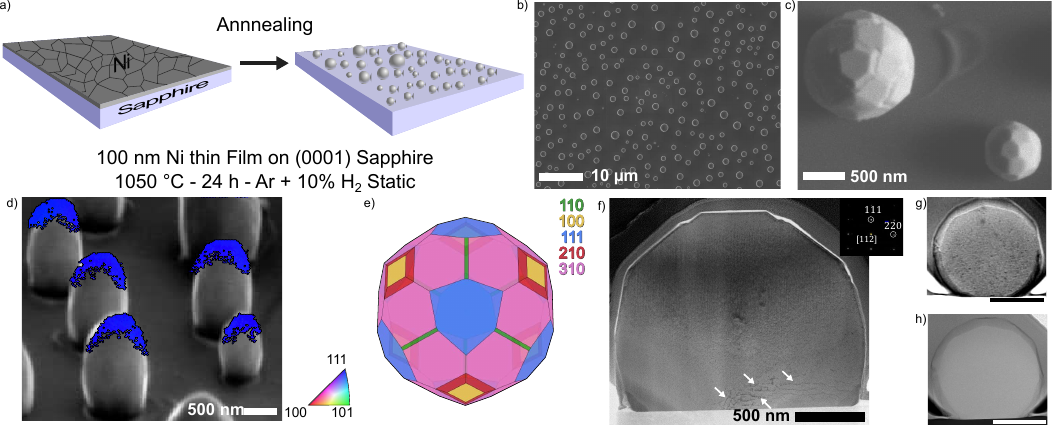}
  \caption{a) Schematic of the SSD process. b) Overview SEM micrograph of the nickel-dewetted microparticles. c) Zoomed-in SEM image of two dewetted microparticles. Left: equatorial diameter $d_e=1600$~nm and top diameter $d_t=480$~nm. Right: $d_e=860$~nm and $d_t=290$~nm. d) EBSD image of the nickel microparticles (inclined view by 70°). Note that only the top facet, which is perfectly oriented with the EBSD detector, can be indexed. e) Wulff construction of the microparticle using Wulffpack in Python \cite{Rahm2020}. f) STEM BF image of a microparticle of the same size as left in c). White arrows indicate dislocations. g) and h) STEM BF and DF image of a microparticle of the same size as the right particle in c). The scale bar is 500 nm. The STEM images have been taken in [112] zone axis.}
  \label{fig:2}
\end{figure}

We obtained a uniform distribution of nickel microparticles of varying sizes by dewetting a 100 nm nickel thin film deposited on sapphire (0001), as illustrated in \blue{\textbf{Figure \ref{fig:2}.a and b}}. \blue{Figure \ref{fig:2}.c} displays two representative microparticles of different sizes at higher magnification. All the particles have the shape of a regular polyhedron truncated below the symmetry center in the middle of the equatorial section. They have self-similar shapes with the same facets and almost identical aspect ratios  (equatorial diameter/height = 1.13 $\pm$ 0.1). Electron backscattered diffraction (EBSD) scans revealed that all the microparticles have the same (111) orientation, as seen in \blue{Figure \ref{fig:2}.d}, which is consistent with previous studies \cite{Sharma2018,Meltzman2011,Hong2009}. By using information from EBSD (\blue{Figure \ref{fig:2}.d}), scanning electron microscopy (SEM) (\blue{Figure \ref{fig:2}.c}), the Wulff construction has been used to calculate the shape of the particles as shown in \blue{Figure \ref{fig:2}.e} (i.e., $\gamma_{hkl}/d_{hkl}=\mathrm{constant}$ where $\gamma_{hkl}$ is the surface energy of the (hkl) plane and $d_{hkl}$ the distance of the facet to the symmetry center, respectively). The surface energies used for this construction are with respect to the energy of the (111) surface: $\gamma_{110}/\gamma_{111}=1.0099$, $\gamma_{100}/\gamma_{111}=1.0913$, $\gamma_{210}/\gamma_{111}=1.0268$, $\gamma_{310}/\gamma_{111}=0.9921$. The same relative energies were utilized to create the particles in the MD models. The shape of the particle obtained was similar to the ones obtained by Hong et al. \cite{Hong2009} using a vacuum atmosphere, where higher index facets like the \{210\} and \{310\} had lower energy making them more stable than lower index facets like \{100\}, \{110\} and \{111\}. The appearance of unexpected crystal facets, which prevent the crystals from reaching their true equilibrium state, is often caused by impurities, atmospheric reactions, or sluggish kinetics of the equilibration as nucleation takes place easier in higher energy surfaces \cite{Hong2009,Wynblatt2009}. The relatively high impurities in the sputtering target together with the encapsulation of the films for annealing could cause the adsorption of those impurities on the surfaces lowering the energy of the surfaces. This depends on the chemical potential of the foreign species and the orientation of the facet which results in a different sequence of energies than expected \cite{Wynblatt2009}. As it can be observed in \blue{Figure \ref{fig:2}.d,f,g,h} the shape of the particle shape is truncated on the bottom by the substrate. This type of construction is called Winterbottom construction and occurs when a crystal equilibrates on a flat substrate of a different material with consequently different substrate--particle interface energies \cite{WINTERBOTTOM1967,Kaplan2013}. In the MD simulations, the model particles have been cut at the bottom along a (111) plane to reproduce the shapes obtained in the SSD experiments.

\medskip

\medskip

As seen in \blue{Figure \ref{fig:2}.c}, larger particles exhibit steps or grooves on the facets. This phenomenon has been explained in literature as the result of impurities being adsorbed onto the surfaces, which then evaporate during the dewetting process \cite{Chatain2005,Chatain2004}. The reason why steps are only present in the bigger particles might be due to the equilibration time needed for removing those extra facets being short enough for the smaller particles but not for the bigger ones. TEM cross-sections of the particles revealed that the small-sized particles are largely dislocation-free, as can be seen in \blue{Figure \ref{fig:2}.g and h} (change of contrast in the edges is due to change in thickness, see \blue{Figure \ref{fig:S1}}). Given the high homologous temperature (0.7$\mathrm{T_M}$  = 1050 °C) and long annealing times (24 h), such single crystalline and pristine nickel particles were expected. However, in bigger particles a few dislocations, marked with white arrows in \blue{Figure \ref{fig:2}.f}, were identified randomly in some regions of the particle. These dislocations were not evenly distributed throughout the particle and are fairly long, which suggests that they were produced during the particle formation and not caused by TEM sample preparation with FIB \cite{FLANAGAN201983}. The dislocation density in this particle has been calculated to be $\sim 2\cdot10^{13} $ m$^{-2}$. In the zoomed-in DF image a different edge around the particle (\blue{Figure \ref{fig:S2}}) is observed which can be due to changes in the thickness of the edge. EDS measurements in that area show a slight increase of oxygen in the particle with respect to the platinum, which can correspond to the native oxide layer on the particle, but it was not possible to perform a more accurate measurement. In contrast, Sharma et al. \cite{Sharma2018} observed a native oxide layer of 4 nm. The effect that this oxide thickness could have on the mechanical properties of microparticles was discussed in Ref.~\cite{Sharma2018}. When an oxide layer surrounds the particle it can soften the contact between the indenter and the sample and decrease stress concentrations in the corners. This leads to higher stresses needed for dislocation nucleation and a higher possibility of homogeneous nucleation of dislocations. 

\medskip

Using a watershed segmentation algorithm in OpenCV library in Python \cite{opencv_library}, the spacing between the dewetted nickel microparticles and their sizes in approximately 2300 particles (11 images like \blue{Figure \ref{fig:2}.b}) was measured from SEM images. The spacing between particles has been assessed by Delaunay triangulation of the segmented images (\blue{\textbf{Figure \ref{fig:3}.a}}), obtaining an average of 3.8 $\mu \mathrm{m}$ from the centers of particles (\blue{Figure \ref{fig:3}.b}, giving an average space of 3 $\mu \mathrm{m}$ between the center of the particle to the edge of the next one, which provides enough clearance to compress individual particles with a 3 $\mu \mathrm{m}$ diameter flat punch diamond indenter, without touching any neighboring particles. To measure the size of the particles three metrics were used: the equatorial diameter, which is the largest averaged diameter in the particle; the top diameter, which is the diagonal of the hexagonal top facet; and the height. The equatorial diameter ranges from 650 to 4500 nm, the top diameters range from 200 to 1900 nm and the heights range from 800 to 2000 nm, as can be observed in \blue{Figure \ref{fig:3}.c}.  \blue{Figure \ref{fig:3}.d} shows an schematic of the equatorial and the top diameter. During in situ micromechanical testing, owing to the high tilt ($\sim$ 75°) employed for aligning the diamond tip with the sample surface, the equatorial diameter can be obtained precisely by imaging the side surface of the particles, but it is difficult to image the top facet clearly. Given that the stress estimates rely on the accuracy of the top diameter estimation, the top and equatorial diameters of 49 random microparticles in high-resolution SEM images (like \blue{Figure \ref{fig:2}.c} were measured before micromechanical testing with ideal imaging conditions in the SEM (flat on the substrate). The inset in \blue{Figure \ref{fig:3}.c} shows the linear relationship between the top and equatorial diameters of the particles that have been identified. Using this relationship, from the accurately determined equatorial diameters of the particles during micromechanical testing, the top diameters of the particles were obtained. The errors of this estimation have been included into the errors of the stress calculation through error propagation.

\subsection{Compression of the microparticles}

In situ tests were performed at different loading conditions inside an SEM to study the mechanical behavior of the nickel microparticles. Over 240 particles have been tested with an Alemnis nanoindenter and three slightly different setups described in section \ref{subsec:mechanical_char} were used. At room temperature, the rate-dependent properties of the microparticles were captured as a function of strain rate from 0.001 to 1000 $\mathrm{s^{-1}}$. At cryogenic temperature, owing to technical limitations in accurately resolving the microscale forces from the noise introduced due to physical vibrations from the liquid nitrogen pump, the mechanical properties of the nickel microparticles could only be measured at strain rates from 0.1 to 10 $\mathrm{s^{-1}}$. It should be noted that this is the first study on the deformation behavior of single crystalline microparticles under such extreme conditions spanning six orders of strain rate magnitude at room temperature (RT) and 3 orders of strain rate magnitude at 128 K. 

\medskip

\begin{figure}[ht]
  \includegraphics[width=\linewidth]{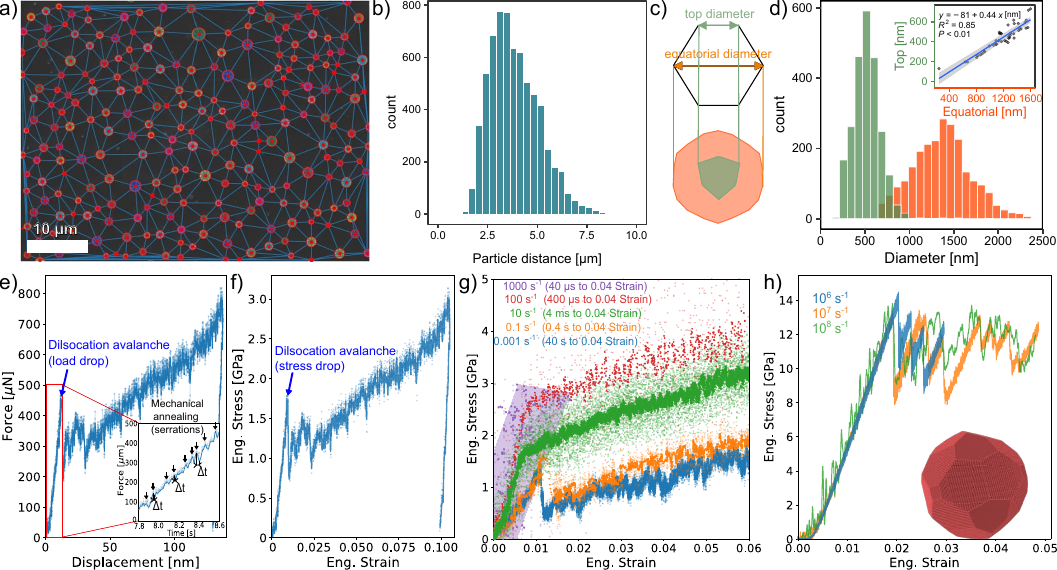}
  \caption{a) Segmentation and Delaunay triangulation of Figure \ref{fig:2}.b. b) Histogram of results of the triangulation to measure the distance between particle centers. c) Schematics of the top and equatorial diameters. d) Histogram of the top (green) and equatorial (orange) diameters of the measured microparticles. The inset shows the linear relationship between the top and the equatorial diameter of the particles, this relation has been used to calculate the top diameter. e) Representative force-displacement curve of a test at 0.01 $\mathrm{s^{-1}}$. The inset shows a zoomed-in force vs. time plot of the same experiment. The arrows indicate the serrations in the curves attributed to mechanical annealing of dislocations. f) Corresponding engineering stress and strain for the same test as in e). g) Representative engineering stress-strain curves of microparticles tested at different strain rates. h) MD engineering stress-strain curves for three strain rates for particles of 40 nm equatorial diameter ($\sim$ 17 nm top facet), inset shows the particle used in the simulations. }
  \label{fig:3}
\end{figure}

\medskip

The corrected force and displacement (\blue{Figure \ref{fig:3}.e}) were then used to calculate the engineering stress and strain (\blue{Figure \ref{fig:3}.f}). The equatorial diameter and height of every particle have been measured before testing. To calculate the strain, the displacement was divided by the height of the particle measured before testing. To calculate the stress, the force was divided by the area of the top facet of the particle (green shade in \blue{Figure \ref{fig:3}.d}). As mentioned before, the top diameter of the particles was estimated with the formula shown in \blue{Figure \ref{fig:3}.c} from the measured equatorial one during testing. This diameter, which is in fact the diagonal of the hexagonal top facet, was used to calculate the top area of this hexagonal top facet, which gave us the upper bound of the stresses, corresponding to the stresses of the top facet under the simplifying assumption of a uniform stress distribution. 

\medskip

The data obtained from the compression tests, i.e. the yield strength, maximum strain and shape of the curves; together with the test parameters like strain rates, temperature and setup utilized and the morphological and size description of the particle has been tabulated. This data was curated, filtered and divided into 4 diameter groups to reduce the influence of the size effects in the rest of the calculations. Due to the insufficient number of data points in the other groups, we have focused the analysis on particles having top facet diameters ranging from 570 to 760 nm for our main calculations (data from 115 particles). Moreover, only groups with more than 3 particles tested were used. Error bars are calculated using error propagation taking into account not only the dispersion of the data points but also the errors due to the calculation of the stresses and the root-mean-square noise of the data at each different strain rate. 

\medskip

\begin{figure}[ht!]
  \includegraphics[width=\linewidth]{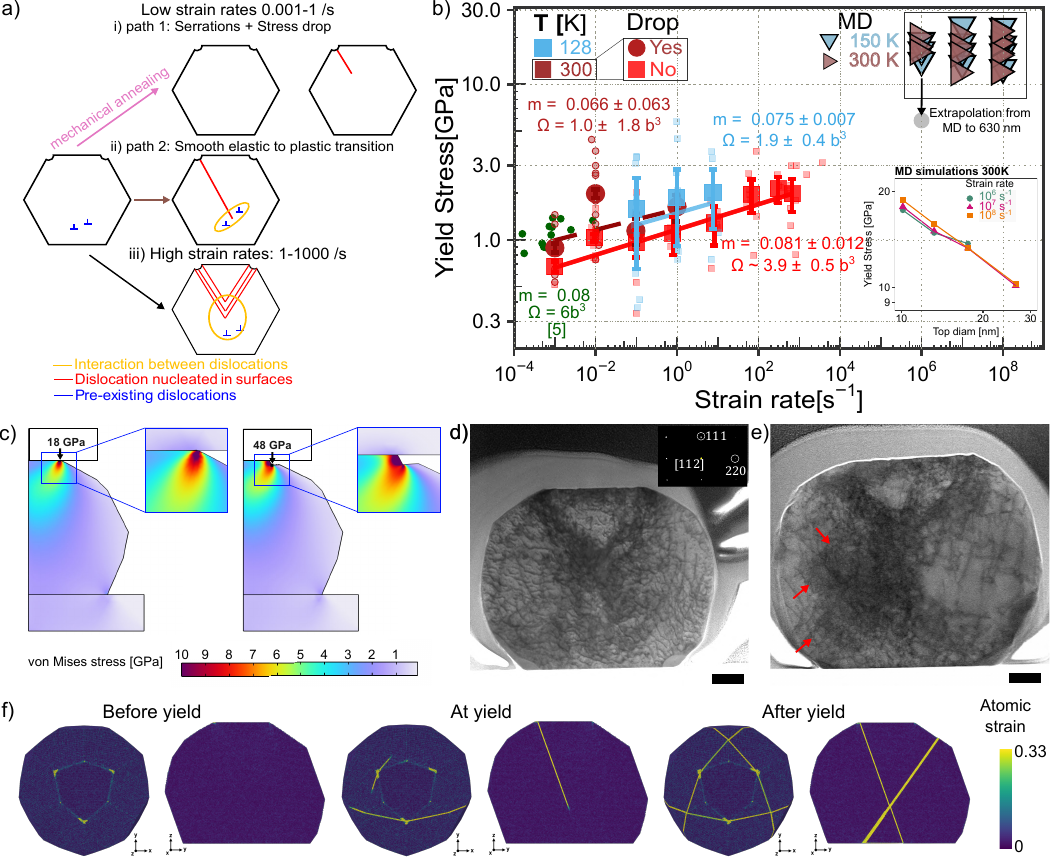}
  \caption{a) Schematics of the proposed deformation mechanisms at low and high strain rates from particles with dislocations.b) Inverse Norton plot for tested particles of 570 to 760 nm top diameter with their corresponding activation parameters: dark red corresponds with the particles showing the stress drop, bright red for the rest of the particles and light blue for the ones tested at 128 K. Data for the MD simulations is shown together with the corresponding extrapolation to bigger sizes (inset shows the size effect shown in the MD simulations at different strain rates). Dark green is the data corresponding to nickel single crystal nanowires with [112] growth direction \cite{Peng2013}.  c) Von Mises stress plots in the 2D FEM simulations of a quarter of the particle with and without the steps on the top facet. d) STEM BF images of particles tested at 0.001 $\mathrm{s^{-1}}$ with a stress drop in the stress-strain signature and e) tested at 1000 $\mathrm{s^{-1}}$, scale bars are 500 nm.  f) Top and central slices of the MD simulated particles color-coded by the atomic strains. The snapshots correspond from left to right to the particle before yield, at yield and after yield, respectively.}
  \label{fig:4}
\end{figure}

\blue{Figure \ref{fig:3}.g} shows the representative stress-strain signatures obtained for strain rates between 0.001 $\mathrm{s^{-1}}$ and 1000 $\mathrm{s^{-1}}$. Note that the dots are the true data points obtained during the test after the corrections described in the Methods section \ref{subsec:data_processing}, i.e. time constant correction, compliance correction and Sneddon correction. Those data points were not subjected to any filtering and the lines correspond to the fitted elastic and plastic portions. In this image, two distinct stress-strain signatures can be observed. At low strain rates up to 1 $\mathrm{s^{-1}}$, a sudden stress (or load) drop after an elastic loading occurs. Conversely, at higher strain rates than 1 $\mathrm{s^{-1}}$, this stress drop no longer occurs and there is a ``smoother'' transition between elastic and plastic deformation. Additionally, at low strain rates ($< 1~\mathrm{s^{-1}}$) several smaller stress drops or serrations occur during the elastic loading of the sample. The inset of \blue{Figure \ref{fig:3}.e} shows the force-time equivalent of the force-displacement segment of the corresponding pillar. These serrations have a timescale dependent on the strain rate. Up to 1 $\mathrm{s^{-1}}$, the serrations ($\sim$ 20 ms) are 2 orders of magnitude smaller than the entire loading segment ($\sim$ 2000 ms). However, at 1 $\mathrm{s^{-1}}$ this difference is reduced to 1 order of magnitude (serrations $\sim$ 1 ms, loading segment $\sim$ 20 ms). It is expected that at even higher strain rates this difference decreases further or even disappears as the serrations could not be resolved at these high strain rates. We believe that at low strain rates (with samples showing stress drop), mechanical annealing takes place, as the dislocation density and mean free path are small enough that they do not interact with each other or produce dislocation sources\cite{meyers1984mechanical, Kiener2006}. Hence, the dislocations present in the particles escape and the stress keeps building up and this yield strength corresponds to the stress required to nucleate dislocations from a surface source \cite{meyers1984mechanical,Kiener2006}. It remains to explain the transition between samples showing a stress drop and the disappearance of this drop at strain rates higher than 1 $\mathrm{s^{-1}}$, seen in \blue{Figure \ref{fig:3}.g}. The sudden drop occurs due to an avalanche of dislocations that nucleates at the weakest points (likely from the defects on the surfaces), see schematics in \textbf{\blue{Figure \ref{fig:4}.a}}, i) low strain rates-path 1. When the strain rates are increased up to $\sim$ 1 $\mathrm{s^{-1}}$, mechanical annealing is less likely to occur as dislocations do not have time to unpin and escape (at 1 $\mathrm{s^{-1}}$, serrations $\sim$ 1 ms, loading segment $\sim$ 20 ms). Given the high stresses promoted by the high strain rates, the dislocations nucleate simultaneously from different nucleation sites e.g. at the surface and start to interact with each other leading to the disappearance of the stress drop, see schematics in \blue{Figure \ref{fig:4}.a}, iii) high strain rates \cite{Ramachandramoorthy2016,Jennings2011}. In other words, at high strain rates the material hardens after yield due to higher dislocation densities, but the yield strength itself is unaffected. Due to the difference in the stress-strain curves, the yield strength has been calculated in two different ways: for the curves at low strain rates where the stress drop is occurring, the maximum stress before the drop is used as the yield point; in the case of the samples without this stress drop the point of intersection between two linear fits of the elastic and the plastic portion, respectively, is used as the yield point (see \blue{Figure \ref{fig:S3}}). Regardless of how it occurs, the yield point keeps increasing monotonically up to 1000 $\mathrm{s^{-1}}$. It is worth mentioning that at low strain rates, some particles do not show this abrupt stress drop (see \blue{Figure \ref{fig:S3}}). In this case, there are two possible explanations: the first is that the alignment between the tip and particle is not perfect, creating artifacts in the loading; the second is that the amount of dislocations in the sample is too high to get annealed out and the dislocations start interacting with each other at low strains resulting in lower yield strengths than their pristine counterparts, see schematics in \blue{Figure \ref{fig:4}.a}, ii) low strain rates-path 2. Mechanical annealing occurring in single-crystal nickel nanopillars have been proved by Song et al. \cite{Song2008}, where they compressed these nanopillars in situ in the TEM. These experiments show that in pillars with high dislocation density $\sim10^{15}$ m$^{-2}$, almost all dislocations can be annealed out during compression. Size plays an important role in this process: it was found that it is easier to anneal smaller pillars ($\sim 150 $ nm in diameter). The monotonic increase in the strength of the particles at higher strain rates is then explained by the increased dislocation flux needed to accommodate the strain rates. Such an increase in dislocation density was observed in literature for sub-µm aluminum samples by in situ straining experiments in the TEM \cite{Oh2009}. This also results in a jump in the apparent hardening seen in the compression after yielding \blue{Figure \ref{fig:2}.h} (for more detailed information please see \blue{Figure \ref{fig:S4}}). Particle compression at lower temperatures shows the same rate-dependent behavior as at room temperature, albeit with an increase in yield strength of around 50 \% (\blue{Figure \ref{fig:4}.b} and \blue{Figure \ref{fig:S5}}). This increase in yield strength with decreasing temperature is similar to the one obtained by Chen et al. \cite{Chen2015}, where they suggest that this dependence is higher than in bulk FCC metals because the surface diffusion is the rate-limiting factor that controls how the material deforms at room temperature or higher temperatures\cite{Sun2014}. However, at lower temperatures, the surface diffusion will not play a significant role and moreover the native oxide layer will hinder surface diffusion. The increase in the strength of the particles with decreasing temperature is due then to the lower thermal energy to overcome the nucleation barrier. 

\medskip

In parallel, molecular dynamics simulations have been performed on particles with the same shape and orientation as the ones obtained in the dewetting experiments, although at smaller length scales (equatorial diameters from 23~nm to 60~nm) to keep the computational demands feasibly low. \blue{Figure \ref{fig:3}.h} shows the stress-strain curves obtained in these simulations (for simplicity, only one size and three strain rates are shown in this graph). A big stress drop can be observed after the nucleation of dislocations at the corners. In contrast, the experiments at high strain rates do not exhibit a big stress drop. This is due to the fact that the particles in the simulations are pristine, without any pre-existing dislocations, and because the sizes of the particles are much smaller than in the experiments. Dislocations in these small particles cannot accumulate and always escape the particle via the surface, bypassing the experimentally observed hardening effect.

\medskip

Comparing our results with the previous compression of nickel microparticles by Sharma et al. \cite{Sharma2018}, the compressive yield stresses obtained in this study are one order of magnitude lower. This can be explained by the fact that their particles have sizes that also are one order of magnitude smaller than the ones tested in this study (see \blue{Figure \ref{fig:1}}). A second reason can be that the facetting of their particles is completely different as the starting film and the annealing conditions for dewetting are different. It has been proven in literature that minimal changes in composition, or annealing conditions can change drastically the facetting of FCC metal particles \cite{Meltzman2011,Hong2009}. That together with the extra facetting steps present in these particles (\blue{Figure \ref{fig:2}.c and f}) can promote earlier yielding in the material. This factor could also explain the relatively high scattering seen in the data, where depending on the roundness of these facetting steps the stress concentrations could be lower or higher \blue{Figure \ref{fig:4}.c} \cite{meyers1984mechanical}. However, if smaller-sized particles are considered, which are more likely to be pristine, the results do not differ greatly and agree with the size effect trend of Sharma et al. \cite{Sharma2018} (see \blue{Figure \ref{fig:S6}}).

\medskip

\medskip

Since there is a change in the stress-strain signatures, it is important to check if the thermal activation parameters can explain this change in deformation behavior. For that, the strain rate dependency of the particles was studied. A clear dependency between the stress-strain response of the microparticles at different strain rates in \blue{Figures \ref{fig:3}.g and \ref{fig:4}.b} was observed. A differentiation between the particles showing a stress drop (for strain rates lower than 10 $\mathrm{s^{-1}}$) and the particles with a smooth transition between elastic and plastic deformation at higher rates was identified. The same tendency was observed for the experiments done at CT. To extract the strain rate sensitivity factor the following equation was used: 

\begin{equation}
    m = \frac{\partial \ln\sigma_y}{\partial \ln\dot{\varepsilon}}
\end{equation}

\medskip
where $\sigma_y$ is the yield strength and $\dot{\varepsilon}$ is the strain rate. At RT, the microparticles showing a sudden stress drop yield a value of $m = 0.066 \pm 0.063$ (number of tests $N = 29$), and at CT $m = 0.075 \pm 0.01$ ($N = 33$). In the case of the microparticles that do not show any stress drop, from 0.001~$\mathrm{s^{-1}}$ to 1000~$\mathrm{s^{-1}}$, the strain rate sensitivity obtained is $m = 0.081 \pm 0.012$ ($N = 53$). Smaller particles (top diameters of 400–570~nm) exhibit nominally smaller strain rate sensitivities (within the error limits): for the particles that do not show stress drop $m = 0.039 \pm 0.066$ ($N = 24$) and for the ones showing drop $m = 0.017 \pm 0.023$ ($N = 42$) (\blue{Figure \ref{fig:S7}}). The similar value of $m$ at both RT and CT and across all the strain rates and also amongst particles showing different stress-strain signatures (i.e. showing a sudden stress drop vs. showing a smoother transition between elastic and plastic regimes) suggest that the same deformation mechanism is responsible. Comparable values of strain rate sensitivity have been previously reported for single crystalline micropillars and nanowires \cite{Peng2013,Xiao2021}. Other studies indicate that single crystalline nickel microcompression renders a lower strain rate sensitivity $\sim 0.001$ in pillars of 3.5~$\mu \mathrm{m}$ diameter \cite{Wehrs2015}. These differences in the strain rate sensitivity can be explained by the difference in the orientation of the samples with respect to the compression axis. Samples with orientations closer to [123] provide a smaller strain sensitivity than samples with an orientation where more than one slip plane is active. Moreover, the increase or appearance of strain rate sensitivity in small-size microcompression experiments was first reported by Jennings et al. \cite{Jennings2011} in copper nanopillars (diameter $\sim$ 80 to 600 nm), where they found $m$ values ranging from 0.027 to 0.057 for the nanopillars, which are five-fold bigger than the ones found for bulk single crystal copper ($\sim 0.006$). All these studies evaluated the strain rate sensitivity up to 0.1 $\mathrm{s^{-1}}$ and showed a dependency of strain rate sensitivity with the size and the orientation of the samples. The studies at high strain rates in nickel have been, to the best of our knowledge, mostly performed in polycrystalline metals, whose strain rate sensitivities are 10-fold smaller than the ones identified in our work for single crystal nickel \cite{Gray1997,Rajaraman2013,Follansbee1991}. Observing no change in the strain rate sensitivity at room temperature in samples not showing a stress drop in comparison with the ones showing the stress drop suggests that only one deformation mechanism is taking place. Another method to investigate the nature of the deformation mechanism is to calculate the apparent activation volume:

\begin{equation}
   \Omega_\text{app} = \sqrt{3} k_B T \frac{\partial \ln \dot{\varepsilon} }{\partial \sigma}
\end{equation}
\medskip

where $k_B$ is the Boltzmann constant and $T$ is the temperature. The activation volume is usually expressed in terms of the Burgers vector of the material, giving a sense of the size scales in terms of atomic volumes involved in the deformation process. With $b = 0.249~\mathrm{nm}$ (partial dislocation $\frac{a}{2}\langle110\rangle$ in FCC), $\Omega_\text{app}$ for nickel at RT and for particles showing a sudden stress drop is $\sim (1.0 \pm 1.8)~b^3$, for the whole range of strain rates and for the particles that do not show a sudden stress drop but a rather smoother transition $\Omega_\text{app} ~ \sim (3.9 \pm 0.5)~b^3$  and for CT $\sim (1.9 \pm 0.4)~b^3$. Those low values typically correspond to surface dislocation nucleation-based deformation ($\Omega_\text{app}  \sim 1-10 ~ b^3$)\cite{Jennings2011}. The smaller activation volume at CT is due to the smaller temperature term and the slightly higher stresses obtained there.  The low values of the activation volume are also in agreement with literature for tests conducted on nickel micropillars and nanowires \cite{Peng2013,Xiao2021}. 

\medskip

The postulated deformation mechanism of the microparticles is that deformation occurs by dislocation nucleation on the weakest points which are the steps around the top facet, since pre-existing dislocations get mechanically annealed out as the compression progresses. To quantify the stress concentration due to these steps, FEM simulations were performed. A 2D approximation of half of a particle was simulated, considering the half as a revolution object. This simulation does not provide the absolute values of the stresses and strains, but it does allow an assessment of the distribution of the stresses. In \blue{Figure \ref{fig:4}.c} snapshots of the FEM simulations at 17 nm of deformation can be seen in particles with and without steps (the provided values on the top being the maximum von Mises stress obtained in the sample). As can be observed, the stresses at the vertex of the steps are around four times bigger than in the smooth particles. Given these results, it can be expected that the weakest points, and hence the nucleation sites in the microparticles in the present study, are the steps on the surface. The size of these steps (tens of nm) is bigger than the size of the oxide layer ($\sim 4 $ nm \cite{Sharma2018}), so the smoothing effect of the oxide reported by Sharma et al. \cite{Sharma2018} is not expected in the current study. Moreover, homogeneous nucleation of dislocations is then expected from these steps, which is an additional theory to support the difference in the strength of the nickel particles between Sharma et al. and this study. TEM analysis of the particles after deformation and MD simulations provide further insights into the deformation mechanisms acting in these particles. 

\medskip

\blue{Figure \ref{fig:4}. d and e} show two particles compressed up to 10\% strain. \blue{Figure \ref{fig:4}. d} corresponds to a test at 0.001 $\mathrm{s^{-1}}$ that shows a stress drop, in contrast with \blue{Figure \ref{fig:4}.e} which corresponds with a test at 1000$\mathrm{s^{-1}}$. It is worth mentioning that the images have been taken in zone axis for better comparison, however, two beam condition images show a homogeneous dislocation distribution in both samples (\blue{Figure \ref{fig:S8}}). In both cases there is a higher dislocation density in the areas corresponding with the (111) planes, however, commenting on the dislocation density in these two particles is difficult as they have been heavily deformed. Nevertheless, there is a significant difference in how the dislocation networks look in both samples. It can be observed in \blue{Figure \ref{fig:4}.e} (particle deformed at 1000 $\mathrm{s^{-1}}$) that dislocation cells have been formed all around the particle (see red arrows in \blue{Figure \ref{fig:4}.e} and \blue{Figure \ref{fig:S9}}), whereas in the case of the particle deformed at quasi-static speeds \blue{Figure \ref{fig:4}.d}, the dislocation networks are not well-defined. This again supports our previous hypothesis, that at high strain rates mechanical annealing is probably not able to occur. We hypothesize that dislocation nucleation at yield consistently starts from the steps around the top facet in all tested particles, regardless of the strain rate. By altering the strain rate, the compressive deformation behavior of nickel microparticles with pre-existing dislocations can be adjusted. At low strain rates, mechanical annealing of these weakly entangled dislocations leads to a stress drop at the yield point due to a dislocation avalanche. In contrast, high strain rates result in a smooth elastic-to-plastic transition at yield. This is attributed to: i) the short time scales (less than 1 ms) of high strain rate experiments preventing mechanical annealing of existing dislocations and ii) the concurrent nucleation of multiple dislocations that interact with each other and with pre-existing ones, causing a higher flow-hardening rate. According to Orowan's theory, higher strain rates necessitate a greater mobile dislocation density to accommodate the strain. Thus, at higher strain rates, a denser network of nucleated dislocations interacts, leading to a smooth elastic-to-plastic transition without a significant stress drop. This enhanced interaction also results in well-defined dislocation cells, as observed in TEM images of particles tested at higher strain rates (\blue{Figure \ref{fig:4}.e} red arrows and \blue{Figure \ref{fig:S9}}).

\medskip

It can be observed in \blue{Figure \ref{fig:4}.b} that MD simulation results show a size effect. This is a counter-intuitive result as dislocations in the simulations nucleate at the corners, where the local stress is only very slightly affected by size. By inspecting the process of dislocation nucleation in more detail, we found that a dislocation loop is pinned by surface features after nucleation (\blue{Figure \ref{fig:S10}}). In order to depin this loop, a stress $\tau \sim 1/R$, where $R$ is the distance between pinning points is needed, as in a typical Frank-Read source \cite{Argon2008}. As the particles are self-similar at different sizes, this distance between pinning points $R$ scales with the size of the particles. With that, a relationship between the yield strength of the particles and their size can be used to model the size effect:

\begin{equation}
    \label{eq:depin}
    \tau = \tau_0 + B/d
\end{equation}
\medskip

where $B$ is a fit constant. Using this relation, the strength of the particles at the MD strain rates can be extrapolated to the sizes of the experimental particles. Obtaining a value of 5.9 GPa extrapolating the data to a particle of a top size of 630 nm (\blue{Figure \ref{fig:S11}}). If the extrapolation is done with the experimental data to a strain rate of 10$^6$ $\mathrm{s^{-1}}$, then the values obtained for the strength of the particles are around 3.8 GPa. Weak surface features and other pre-existing defects can significantly reduce the particles' strength compared to the pristine computer model. Given these fundamental differences between experiments and simulations, for instance, the pre-existing dislocations, we consider the results to be in good agreement. Such dislocation depinning has been observed before in simulations of nanocrystalline metals \cite{Bitzek2008}. The present results of the scaling from Equation~\ref{eq:depin} applying to simulation and experiment thus suggest that dislocation pinning can also explain size effects in certain nanoparticle shapes.  Moreover, in \blue{Figure \ref{fig:4}.f} it can be observed how the microparticles look during deformation in MD simulations. Just before yield the local strain is concentrated on the vertex of the top facet of the particle. After yielding it can be observed how the elastic strain has propagated along the (111) plane directions not only in one of the vertex but already in two (only visible from the top). Finally, after yielding, the deformation in the particle is propagating along the three available (111) planes. It can be observed that the local atomic strains in the simulated particles coincide with the areas where higher dislocation densities are seen experimentally in the TEM images of the tested particles \blue{Figure \ref{fig:4}.d and e}.

\section{Conclusions and outlook}

We report for the first time the properties of single crystalline nickel microparticles at a wide range of strain rates, from 0.001 to 1000 $\mathrm{s^{-1}}$ and at cryogenic temperatures (128 K). Moreover, such extreme micromechanical tests allowed a better comparison of the experimental results against MD simulations and they yield comparable results when taking into account size and rate effects. The following conclusions can be drawn:
\begin{itemize}
    \item The nickel microparticles of sizes ranging from 570 to 760 nm exhibit two different stress-strain behaviors upon compression up to strain rates of 1 $\mathrm{s^{-1}}$. In one case, during loading, mechanical annealing of pre-existing dislocations can take place and the stress keeps building up until a sudden stress drop occurs. In the other case, the pre-existing dislocations cannot escape the particle (possibly because they were unable to depin or because the dislocation density on the particles was too high). In this case, the yield strength was slightly lower and a smoother transition between the elastic and plastic regimes was observed, as the dislocations that nucleate start to interact with the preexisting dislocations. 
    \item Regardless of these differences in the shapes of the stress-strain signatures, and based on thermal activation analysis, we believe that the deformation mechanism taking place in the microparticles is surface dislocation nucleation. This is supported by the very small activation volume of $< 3b^3$.
    \item As the strain rate is increased to values beyond 1 $\mathrm{s^{-1}}$, not all compression experiments show a sudden stress drop and the yield strength increases monotonically with the strain rate. The monotonic strength increase is attributed to the increase of dislocation flux needed to maintain the strain rate. In this case, we postulate that the dislocations do not have sufficient time to unpin and get annihilated at the surfaces of the particles, again leading to a smoother transition from elasticity to plasticity. 
    \item Lower temperatures, here 128 K, also do not change the deformation mechanisms of Sx nickel microparticles. They show a simply slightly higher strength than their room temperature counterparts, which is expected given the lower thermal activation energy.
    \item By conducting compression experiments on  single crystalline dewetted microparticles with sizes in the range of hundreds of nanometers at strain rates up to 1000 $\mathrm{s^{-1}}$, we can go a step further in closing the gap between MD simulations and experiments.
    
\end{itemize}

\medskip

New developments in testing platforms like microelectromechanical systems and advancements in microelectronics can assist in closing this gap between simulations and experiments even further. One of the main limitations of the present setup is the increase of noise at higher sampling frequencies that are required as the strain rate increases, making the signal-to-noise ratio insufficient for testing even smaller particles ($\sim  < $ 450 nm top diameter) at high strain rates ($>$ 100 $\mathrm{s^{-1}}$). Developing new approaches that improve the signal-to-noise ratio will enable the dynamic testing of these smaller particles at strain rates beyond 10000 $\mathrm{s^{-1}}$. Moreover, more stable cryogenic setups that reduce the vibrations will be desirable to again improve the signal-to-noise ratio and improve the quality of the results. These advancements will enable a 1 to 1 comparison with MD simulations in future. Moreover, the extension of this study to different temperatures, e.g. liquid helium or high temperatures, will complete the deformation map of nickel and other similar FCC metals further.

\section{Materials and Methods}\label{sec:methods}
\subsection{Dewetting experiments}
100 nm thick nickel films of purity 99.95\% were sputter deposited on $\alpha$-Al$_2$O$_3$ (0001) wafers (miscut $<$0.1°, one-side polished, thickness 0.33 mm, CrysTec GmbH). Common impurities in Ni are cobalt, copper, iron or sulfur \cite{wilhelm1963high}. Prior to the deposition, the base pressure of $5\cdot10^{-8}$ mbar was attained in the deposition chamber. The sputtering was carried out at a power of 100 W in Ar atmosphere at the pressure of $5\cdot 10^{-3}$ mbar. The substrate was kept at room temperature during film growth. 

\medskip

Subsequently, the film with the substrate was cut into smaller pieces of about $5 \times 5$~mm, which were used for annealing. The samples were encapsulated in silica ampules filled with a mixture of Ar $+$ 10\% H$_2$ at 300 mbar. Before the encapsulation, the atmosphere was purged with pure Ar two times. The encapsulated samples were annealed at homologous temperature $0.7 T_M$ (1050 $^{\circ}$C, below nickel melting point 1455 $^{\circ}$C) for 24 h. The annealing was carried out in a conventional high-temperature muffle furnace.

\subsection{Microstructural and compositional characterization of the film and particles}
The composition of the thin film before annealing was examined using a Rigaku SmartLab Diffractometer (XRD) with a copper rotating anode ($\lambda$ K$\alpha$1=0.15406 nm), a Goebel mirror for parallel-beam geometry, and an energy-dispersive 2D detector. Since the films remained on the substrate, conventional $\theta$–2$\theta$ scans were conducted in reflection mode over a 2$\theta$ range of 20-120° with a step size of 0.01° and a scan speed of 4° min$\mathrm{^{-1}}$ over an area of 500 × 500 $\mathrm{\mu m^2}$. The collected patterns were analyzed using DIFFRAC.EVA version 4.3.0.1, along with a structural model containing the crystallographic information of nickel and Al$_2$O$_3$. These analyses were performed to verify the composition and orientation and to rule out any potential contamination. \blue{Figure \ref{fig:S12}} shows the diffraction data for nickel and sapphire, no peaks for other materials could be identified.

\medskip

The equilibrated nickel particles were examined in the SEM to prove their viability to be tested and analyze their size and distribution. Additionally, EBSD and electron dispersive X-ray (EDX) were used to confirm their composition and orientation and to again discount any contamination that could have occurred during the handling and encapsulation of the film. To avoid charging during microstructural and mechanical characterization in the SEM the samples were coated with a thin layer of carbon (2-4 nm) with a Precision Etching and Coating System (PECS), Gatan, equipped with a thickness sensor. 

\medskip

Further microstructural characterization was carried out by SEM and TEM on the particles prior to compression and after compression. The cross-sectional specimens were prepared with a ThermoFisher Scios 2 dual beam FIB/SEM using the lift-out technique. To protect the particles from ion damage, first Pt was deposited with the electron beam, subsequently, a thicker layer was deposited using the ion beam Pt-assisted growth. 1-2 $\mu \mathrm{m}$ lamellae were cut out of the films and transferred to a TEM grid. Finally, the lamellae were thinned down using 30 kV and currents from 1 nA down to 0.3 nA. To minimize the beam damage, the final thinning of the lamellae was carried out at 5 kV.

\medskip

The TEM characterization of all the particles was performed on a probe-corrected FEI Titan Themis 80-300 (Thermo Fisher Scientific) S/TEM and JEOL JEM 2100Plus operated at 300 and 200 kV, respectively. Diffraction analysis of the particles was performed in a JEOL JEM 2100Plus. The sapphire and nickel particles were oriented in a low-index zone axis by Kikuchi electron diffraction. After the alignment of a nickel particle, selected area electron diffraction patterns were acquired to identify the facet planes. The two beam condition was used to acquire the DF and BF TEM and STEM images using JEOL JEM 2100Plus microscope.

\medskip

The orientation of the top facet of the particle was ascertained by EBSD. The EBSD has been taken with a step size of 30 nm to be able to resolve the upper facet. Subsequently, using the diffraction patterns, the planes corresponding to the facets were identified. Finally, the experimentally determined surface anisotropy and Wulffpack Python library \cite{Rahm2020} were used to simulate the shape of the faceted particles.

\medskip

Dislocation density was qualitatively measured from the TEM images of the particles by calculating the total length of dislocations line per unit volume of particles \cite{Ham1961}: 

\begin{equation}
    \rho \approx \frac{L_t}{A ~ t} 
\end{equation}

where $L_t$ is the total length of the dislocations, $A$ is in this case the area of the particle, and $t$ is the thickness of the lamellae. To calculate the total length of the dislocation segments the images have been manually annotated, after that the images have been segmented using ImageJ and the length of the dislocation segments has been calculated using the Skeletonize plugin \cite{skeletonize_plugin}.

\subsection{Mechanical characterization}\label{subsec:mechanical_char}

Compression of the microparticles was carried out using a circular diamond flat punch of 3 $\mu \mathrm{m}$ in diameter (Synton-MDP AG, Switzerland) using a nanoindenter (Alemnis AG, Switzerland) installed in a Zeiss Gemini 500 SEM (Zeiss, Germany). Before testing, the equatorial diameter and height of each particle was measured individually. The compression experiments have been performed under displacement control at room temperature and cryogenic temperatures. In the case of room temperature, the test has been performed across six orders of magnitude in strain rates from 10$^{-3}$ $\mathrm{s^{-1}}$ up to 10$^3$ $\mathrm{s^{-1}}$. For the test at low strain rates, i.e., $<$10 $\mathrm{s^{-1}}$, the test was carried out using the Alemnis Standard Assembly (ASA) equipped with a strain gauge load cell that is limited to 10 $\mu \mathrm{m}\mathrm{s}^{-1}$ due to the load cell's susceptibility to ringing at even faster speeds of actuation. 
In order to achieve strain rates beyond 10 $\mathrm{s^{-1}}$, a piezo-based load cell, together with a piezo tube actuator, was utilized instead of the traditional strain gauge load cell, and the piezo stack-based actuator. This allowed for a maximum speed of 10 mm$\mathrm{s^{-1}}$ without the interference of inertia or resonances. Due to the short duration of the experiments, the load and displacement data were acquired using high-acquisition frequency oscilloscopes with capabilities of up to 5 GHz. 

\medskip

The tests at cryogenic temperatures have been performed from 0.1 $\mathrm{s^{-1}}$ up to 10 $\mathrm{s^{-1}}$. The LTM-Cryo system comprises a cold finger cooled with liquid N$_2$ pumped from an external dewar into the system. Copper braids link the cold finger to the tip and the sample, while ceramic shafts isolate it thermally from the indenter frame. Tip, sample, and frame have individual resistive heaters and thermocouples, which work in a closed feedback loop to allow full control of the system's temperature. To minimize drift, the frame of the testing system is maintained at a constant temperature using the frame heaters and temperature feedback. After cooling down the system, the frame heater are turned on and the frame is allowed to stabilize until the change in frame temperature is less than $\pm$ 0.005°C in 10 minutes. Prior to performing actual compression on the particles, temperature matching between the indenter and the tip is performed to minimize drift. For this, several indentations under load control were performed, where the indenter or sample temperature with respect to the other are changed by $\pm$ 5°C. Then the drift is measured in a hold segment of 100 s during the unload. The pair of temperatures that provided a drift lower than 200 pm $\mathrm{s^{-1}}$ were chosen. This procedure is repeated at the start of every set of tests.

\subsection{Data processing}\label{subsec:data_processing}
 Exhaustive data processing has been applied to extract the data from high-speed and cryogenic experiments. For that, Pandas \cite{mckinney-proc-scipy-2010,reback2020pandas}, SciPy \cite{2020SciPy-NMeth} libraries in Python have been used. Given the complexities of the setups, different strategies have to be applied and three different corrections have to be carried out. Firstly, in the case of tests at high strain rates, where oscilloscopes record the data, given the different electronics involved, the signals for the displacement and the load have to be synchronized \cite{Lalith2024}, for a calibration of the electronics-dependent lag that takes to increase the load when the displacement increases have been carried out elsewhere \cite{Lalith2024}. Once the two signals are synchronized, a time constant correction, to account for the time that the sensors take to respond to fast events is performed \cite{Merle2019}. It is known that sensors do not have an immediate instantaneous response and this becomes an important factor when the strain rates are in the range of 1000 $\mathrm{s^{-1}}$. Finally, zero-phase low-pass filters are used to filter higher-frequency ($\sim$ 600 kHz, see \blue{Figure \ref{fig:S13}}) noises in the case of the test at high strain rates. In the case of the tests at cryogenic temperatures, the vibrations caused in the system by the liquid N$_2$ being pumped into the frame have been characterized (35, 60, 80, 120, 170, 230, 360, 500 Hz, see \blue{Figure \ref{fig:S14}}) and a bandstop zero phase filter 3\% above and below those frequencies has been used to minimize the contribution of these vibrations.

The data has been analyzed using Python and R. The displacement data has been corrected for the frame stiffness taking into account the different setups used by:

 \begin{equation}
     d_\text{compliance} = d\cdot C_\text{frame}
 \end{equation}
 
where $d$ is the measured displacement and $C_\text{frame}$ is the compliance of the frame. After that, the displacement has been corrected for the sink in the substrate with the Sneddon correction. 

\begin{equation}
    d_\text{Sneddon}=d_\text{compliance}-C_\text{Sneddon}\cdot F
\end{equation}

where $F$ is the instantaneous force measured by the system and $C_\text{Sneddon}$ is the compliance caused by the elastic sink-in into the substrate and is given by: 

\begin{equation}
    C_\text{Sneddon} = \frac{(1-\nu^2)\sqrt{\pi}}{2\cdot E\sqrt{A_p}}
\end{equation}

where $\nu$ and $E$ are the Poisson ratio and the Elastic modulus of the sapphire, respectively; and $A_p$ is the cross-sectional area of the particle. In this case, the area of the hexagonal top facet in \blue{Figure \ref{fig:3}.d}) was used. The corrected force and displacement of a microparticle compressed at 0.01 $\mathrm{s^{-1}}$ can be observed in \blue{Figure \ref{fig:3}.e)}.

\subsection{Finite Element Simulations}

FEM simulation on the mechanical behavior of the particles were performed using COMSOL Multiphysics. Two different models were considered: an ideal particle without and without steps. For simplicity and speed of the calculations, 2D models of the particles were used, considering the left edge as a revolution axis. The model was meshed automatically with COMSOL free triangular elements and refined in the corners where stress concentrations were expected. The substrate and indenter have been modeled as rigid materials. The contact of the particle with the substrate has been modeled as a fully coupled contact pair and the contact with the indenter was modeled according to a Coulomb friction, with a friction coefficient of 0.1\cite{Soler2012}. The base of the supporting material was fully constrained. A stationary study was used to simulate the compression where the indenter was assigned by an incremental displacement to be able to capture different snapshots of the deformation and aid the convergence of the system. The material behavior of nickel in the (111) direction was modeled using a linear elastic material model with the Voigt notation.   

\subsection{MD Simulations}

MD simulations were performed using LAMMPS \cite{Plimpton1995, Thompson2022} with an EAM potential for nickel by Mishin et al. \cite{Mishin2004} and a time integration step of 2 fs. The nanoparticles of different shapes (see below) were equilibrated at target temperatures of either 150 or 300 K using a Nos\'e-Hoover thermostat for 200 ps. For each particle shape and temperature, 3 independent equilibration runs with different velocity initializations were performed. Then the particles were compressed at a fixed true strain rate by two moving, harmonic walls. These walls apply a repulsive spring force $F = 2k r$ per atom that has passed through the wall plane, where $r$ is the shortest distance between atom and wall and $k$ is the spring constant. This spring constant can be regarded as the local contact stiffness between particle and wall. The reaction forces at the walls were recorded and converted to a stress by dividing by the contact area at the top of the particle in order to use the exact same methodology as in the experiment. We tested values of $k = 1$, 10, 100, $1000~\mathrm{eV \r{A}^{-2}} = 0.016$, 0.160, 1.602, $16.022~\mathrm{N mm^{-1}}$ on an example particle. We could not find significant differences in Young's modulus and yield stress of the particle beyond the statistical fluctuations, except for a small deviation at $k = 1~\mathrm{eV \r{A}^{-2}}$ \blue{(Figure \ref{fig:S15})}. We thus used $k = 100~\mathrm{eV \r{A}^{-2}}$ for the rest of the simulations. True strain rates of $10^6$, $10^7$, and $10^8~\mathrm{s^{-1}}$ were used.

\medskip

Particles in Wulff shapes were created using the Atomic Simulation Environment (ASE) \cite{Larsen2017} with diameters of the equatorial shape of 23, 30, 40, and 60~nm (around 500\,000 to 9\,000\,000 atoms) using the calculated relative energies for the experimental particles. For the largest particle, only one independent run was performed per strain rate and temperature to save computing time, since we did not find large fluctuations of yield strength between independent runs for smaller diameters. We also excluded the strain rate $10^6$/s for the largest particle, since the strain rate effects were weaker than the size effects. In the experiments, the bottom surface facet of the particle is bigger than the top facet due to the substrate-particle interface energy \cite{WINTERBOTTOM1967}. We matched that shape by cutting the particle below a (111) plane positioned to match the approximate experimental Winterbottom particle shape. Compression simulations at strain rates of $10^8$/s revealed no differences in the reaction force and yield stress between the different shapes, although the yield strain and apparent stiffness varied \blue{(Figure \ref{fig:S16})}. We used the more realistic Winterbottom shape for further simulations.

\medskip

Yield stress was extracted by applying a moving average filter on the stress-strain data to remove thermal fluctuations and then taking the maximum value. Stress-strain curves with extracted yield stresses marked are shown in \blue{Figure \ref{fig:S17}} for one independent run per particle.

\medskip
\textbf{Supporting Information} \par 
Supporting Information is available from the Wiley Online Library or from the author.

\medskip
\textbf{Acknowledgements} \par 
Benjamin Breitbach is acknowledged for XRD measurements. We thank Dr.~Hanna Bishara for the deposition of the nickel film. We thank Prof.~Erik Bitzek for useful comments and discussions. B.B.\ \& D.S. were supported by the Alexander von Humboldt Foundation. B.B. was supported by the Marie Sk\l{}odowska-Curie individual fellowship (Grant agreement No. 101064660; DyThM-FCC). T.B.\ \& G.D.\ acknowledge funding from the European Research Council (ERC) under the European Union's Horizon 2020 research and innovation program (Grant Agreement No.~787446; GB-CORRELATE). R.R.\ would like to acknowledge funding from the ERC (Grant agreement No. 101078619; AMMicro) and Eurostars Project HINT (01QE2146C). Views and opinions expressed are however those of the author(s) only and do not necessarily reflect those of the European Union or the European Research Council. Neither the European Union nor the granting authority can be held responsible for them. 

\medskip
\textbf{Authorship contribution statement}

\textbf{Bárbara Bellón:} Conceptualization, Investigation - dewetting, compression experiments, high strain rate experiments, TEM, Data Analysis, Writing - original draft, review \& editing. 

\textbf{Lalith Bhaskar:} Investigation - high strain rate experiments, Writing - review \& editing. 

\textbf{Tobias Brink:} Conceptualization, Investigation - molecular dynamics simulations, Writing - review \& editing. 

\textbf{Raquel Aymerich-Armengol:} Investigation - TEM, Writing - review \& editing. 

\textbf{Dipali Sonawane:} Investigation - dewetting, Writing - review \& editing. 

\textbf{Dominique Chatain:} Conceptualization, Writing - review \& editing

\textbf{Gerhard Dehm:} Conceptualization, Supervision, Writing - review \& editing. 

\textbf{Rajaprakash Ramachandramoorthy:} Conceptualization, Supervision, Writing - original draft, review \& editing.

\medskip

\bibliographystyle{ieeetr}
\bibliography{bibliography}

\newpage
\centering
\title{\textbf{\LARGE{Supporting Information}\\
\Large{High strain rate compressive deformation behavior of Nickel Microparticles}}}

\medskip
\author{Bárbara Bellón$^{1*}$, Lalith Kumar Bhaskar$^{1}$, Tobias Brink$^{1}$, \\Raquel Aymerich-Armengol$^{1,2}$, Dipali Sonawane$^{1}$, Dominique Chatain$^{3}$,\\ Gerhard Dehm$^{1}$, Rajaprakash Ramachandramoorthy$^{1*}$}
\medskip

\centering{
\small{$^1$ Max Planck Institute for Sustainable Materials, 40237 Düsseldorf, Germany\\
$^2$ Current address: Center for Visualizing Catalytic Processes, Department of Physics, Technical University of Denmark, 2800 Kgs. Lyngby, Denmark\\
$^3$ Aix-Marseille Univ. CNRS, CINAM, Marseille 13288, France
}}
\setcounter{figure}{0} 
\renewcommand{\thefigure}{S\arabic{figure}}
\section*{Supplementary Figures}
\begin{figure}[h!]

\centering
  \includegraphics[width=0.5\linewidth]{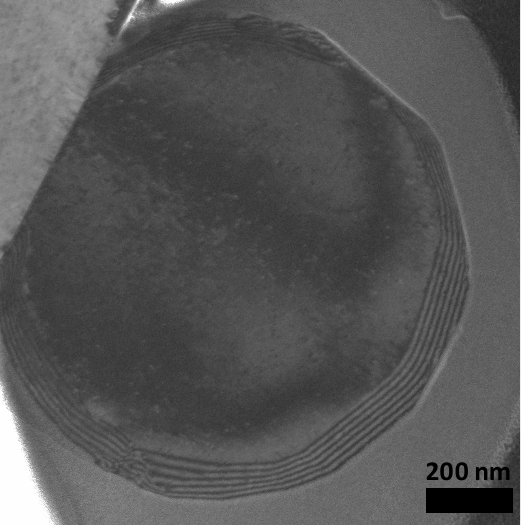}
  \caption{DF TEM of particle Figure 2.g) and h) showing thickness fringes around the particle and supporting that the change of contrast seen in the STEM pictures is due to change of thickness.}
  \label{fig:S1}
\end{figure}

\begin{figure}[h!]
  \includegraphics[width=0.9\linewidth]{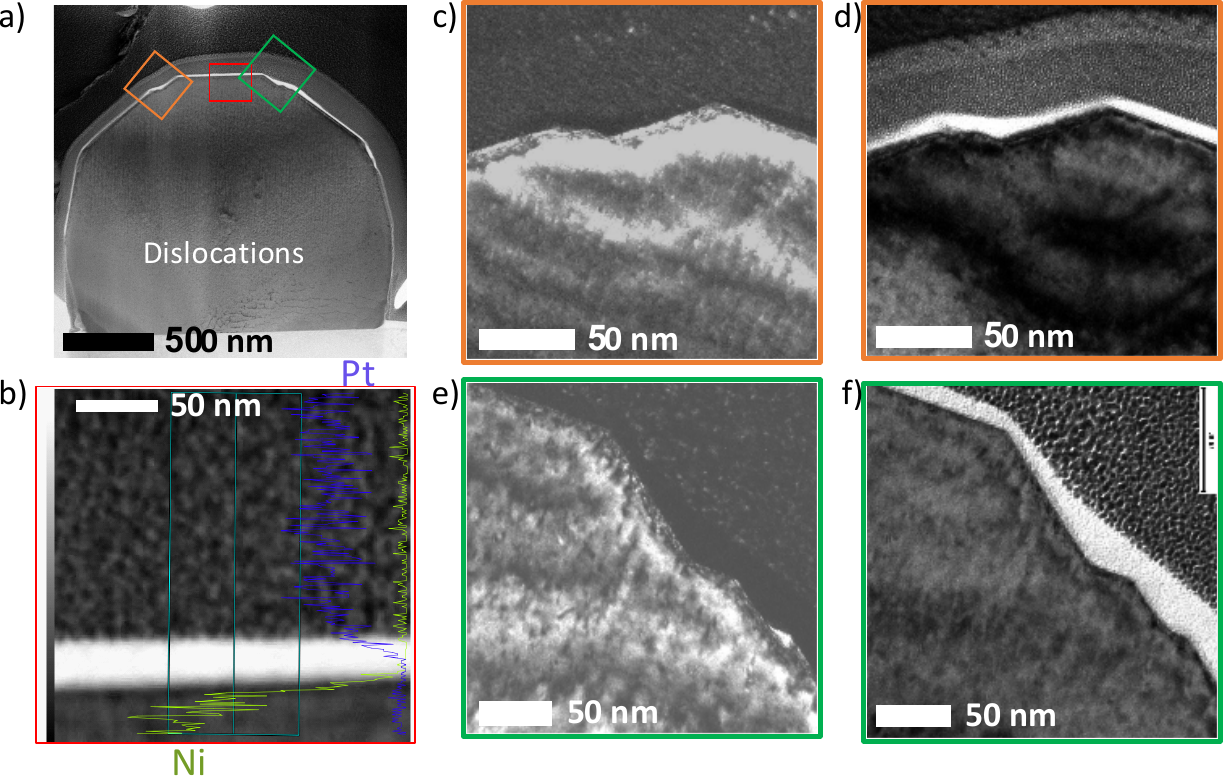}
  \caption{a) Same STEM image of Figure 2.f). b) Increased magnification of the central top part of the particle with a superposed EDX profile over the light blue box. Pt is shown in dark blue and Ni in bright green; c) DF and d) BF TEM zoomed orange square in a); e) DF and f) BF TEM zoomed green square in a).}
  \label{fig:S2}
\end{figure}

\begin{figure}[h!]
\centering
  \includegraphics[width=0.6\linewidth]{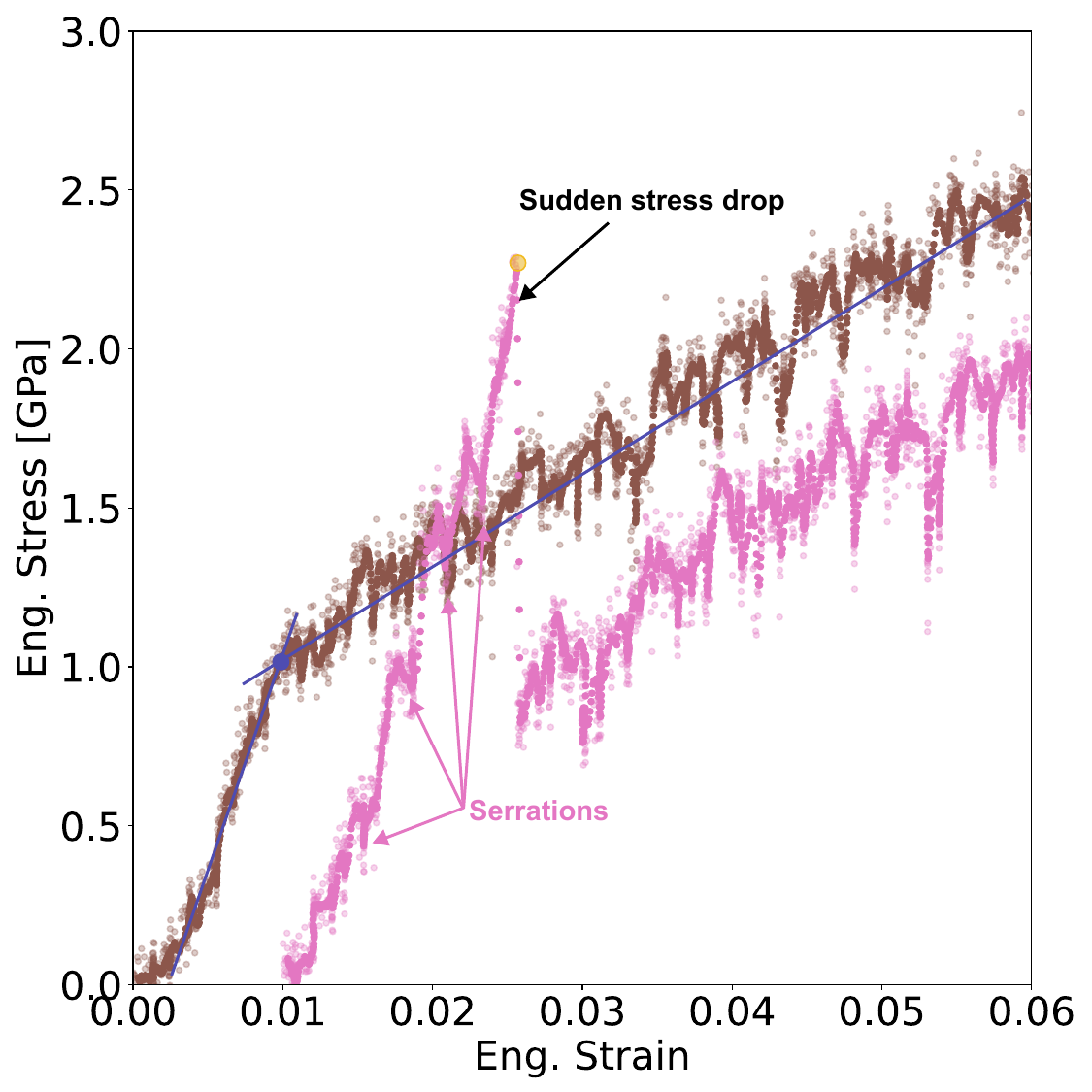}
  \caption{Engineering stress-strain curves of two microparticles of similar sizes (top facets 610 nm brown and 550 nm pink). The brown curve shows a smooth transition between elasticity and plasticity, characterized by the yield strength. The pink curve shows a serrated elastic loading followed by a sudden stress drop. We believe that the serrations corresponds to dislocations getting annealed out in the surface o the particle. When it is not possible to accommodate more deformation, there is an avalanche of dislocations nucleating from a possible weak link in the surface. The yield strength on particles without the sudden load drop are calculated with the intersection (blue dot) between the fitting of the elastic and the plastic portions of the deformation (blue lines). In the case of the particles showing the sudden drop the yield strength has been calculated as the maximum stress before the drop (yellow dot).}
  \label{fig:S3}
\end{figure}

\begin{figure}[h!]
\centering
  \includegraphics[width=0.5\linewidth]{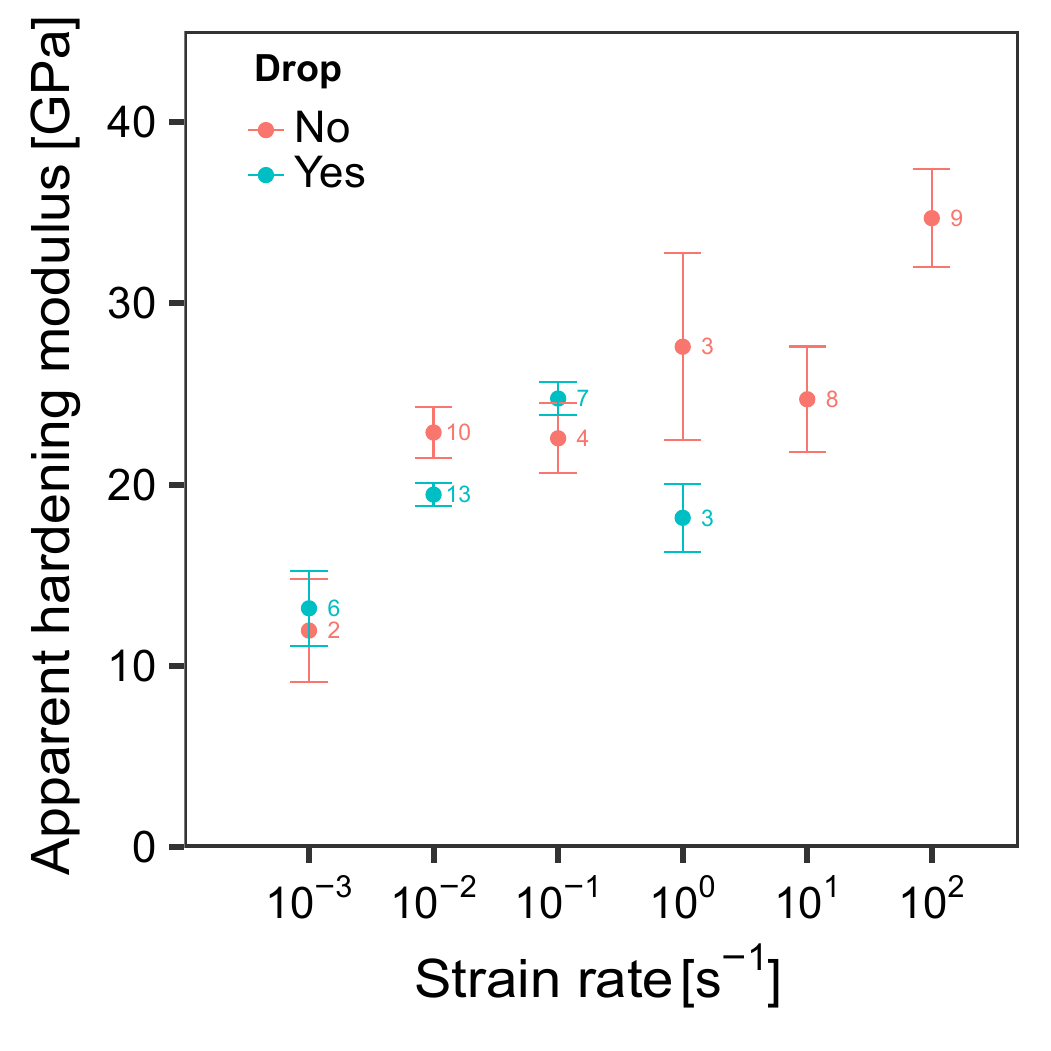}
  \caption{Apparent hardening in GPa vs. strain rate in $\mathrm{s^{-1}}$ after yielding. Particles showing a sudden stress drop are depicted in light blue and a smooth transition (or No drop) are depicted in red. At higher strain rates $>10~\mathrm{s^{-1}}$ there is an increase of the apparent hardening attributed to the increase of dislocation flux necessary to maintain the strain rates. Error bars are the standard deviation of the apparent hardening modulus measured in individual particles. The numbers to the right of the images are the number of particles in which the apparent hardening modulus has been measured}
  \label{fig:S4}
\end{figure}

\begin{figure}[h!]
  \includegraphics[width=\linewidth]{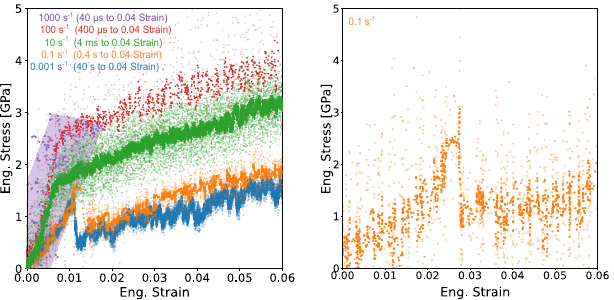}
  \caption{Representative engineering stress-strain curves at room temperature (left) and cryogenic temperature (128 K)}
  \label{fig:S5}
\end{figure}

\begin{figure}[h!]
\centering
  \includegraphics[width=0.7\linewidth]{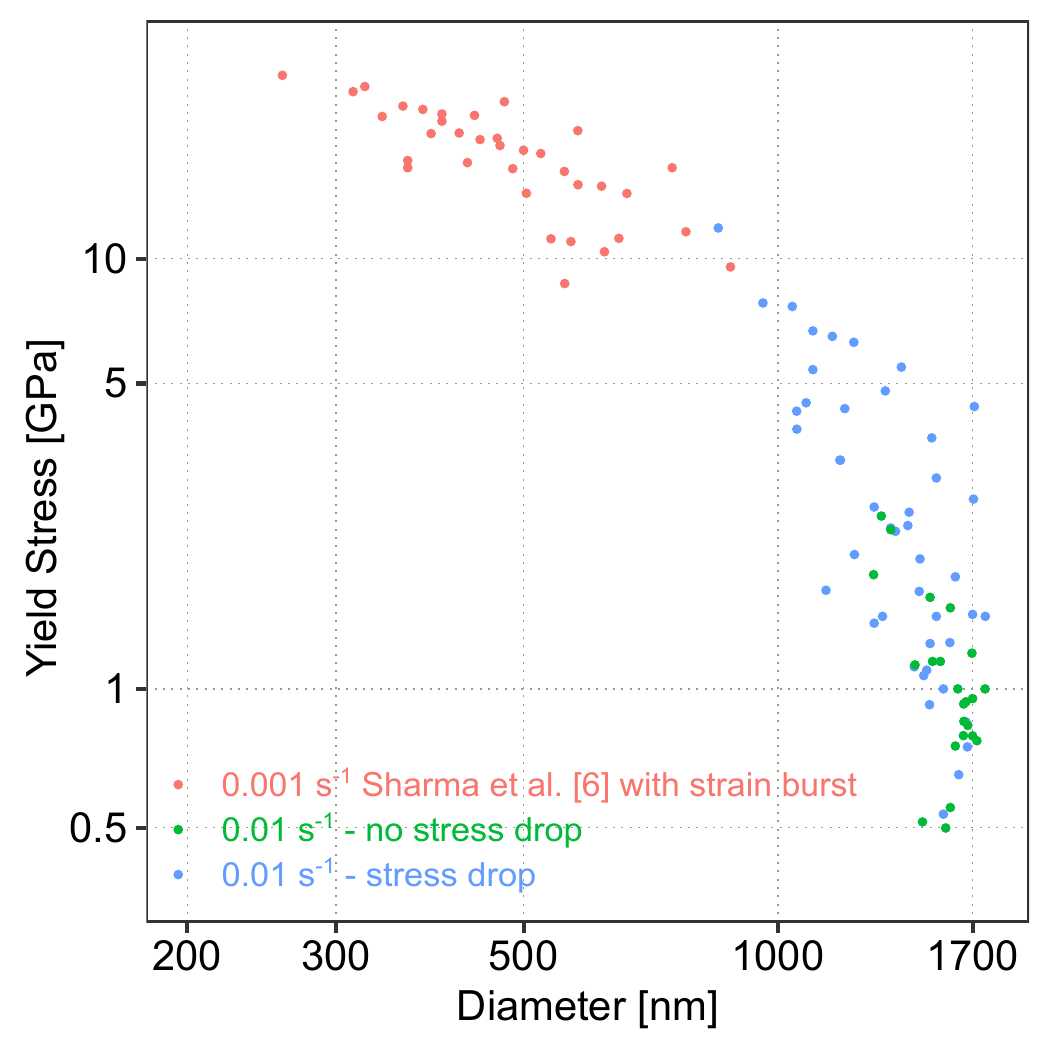}
  \caption{Size effect plot (yield strength in GPa vs diameter of the top facet in nm) of the particles compressed by Sharma et al. [6] in blue, and the ones compressed at low strain rates 0.01 $\mathrm{s^{-1}}$ in this work. The particles showing sudden stress drop are depicted in red and the ones without are depicted in green.}
  \label{fig:S6}
\end{figure}

\begin{figure}[h!]
\centering
  \includegraphics[width=0.8\linewidth]{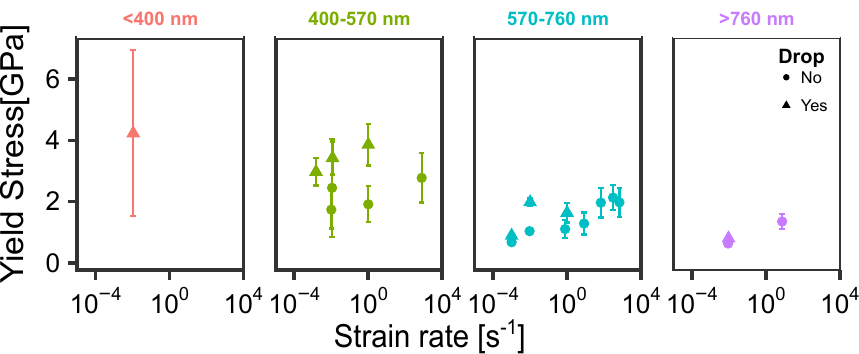}
  \caption{Yield strength in GPa vs strain rate in $\mathrm{s^{-1}}$ of the 4 diameter groups mentioned in the main text for the test at room temperature. Lack of statistics prevents calculating accurately the thermal activation parameters in particles with top diameters out the range of 570 to 760 nm. The tests have been stratified between particles showing a stress drop (depicted with triangles) and particles not showing a stress drop (depicted with circles). The error bars have been calculated using error propagation. A size effect can be observed in this graph. }
  \label{fig:S7}
\end{figure}

\begin{figure}[h!]
\centering
  \includegraphics[width=0.9\linewidth]{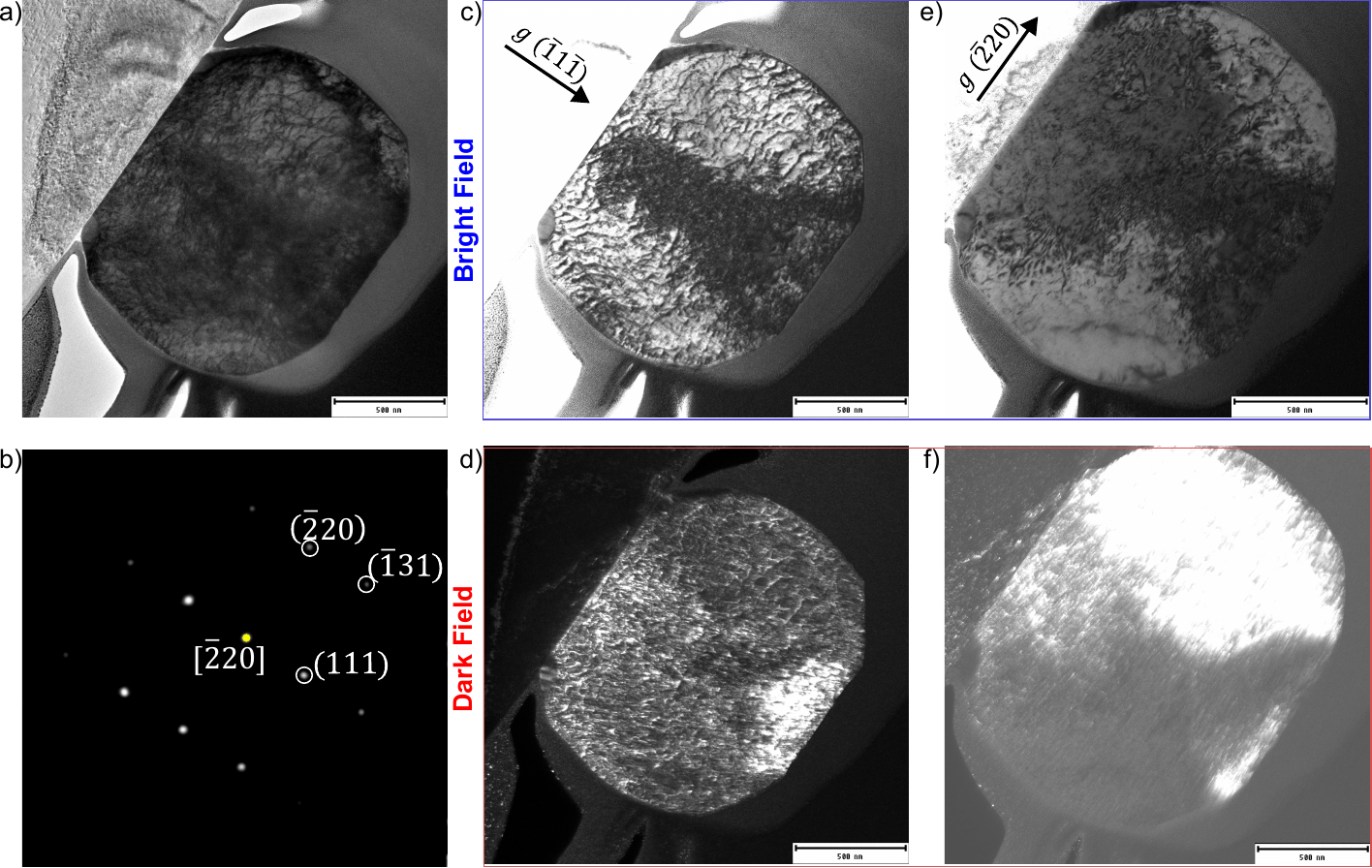}
  \caption{a) TEM of a particle deformed at 0.001 $\mathrm{s^{-1}}$ in zone axis. b) Indexed diffraction pattern of the particle in Figure \ref{fig:S8}.a). c) BF image and d) DF image of the particle taken in two beam conditions with g = $\bar{1}1\bar{1}$. e) BF image and f) DF image of the particle taken in two beam conditions with g = $\bar{2}20$}
  \label{fig:S8}
\end{figure}

\begin{figure}[h!]
\centering
  \includegraphics[width=0.9\linewidth]{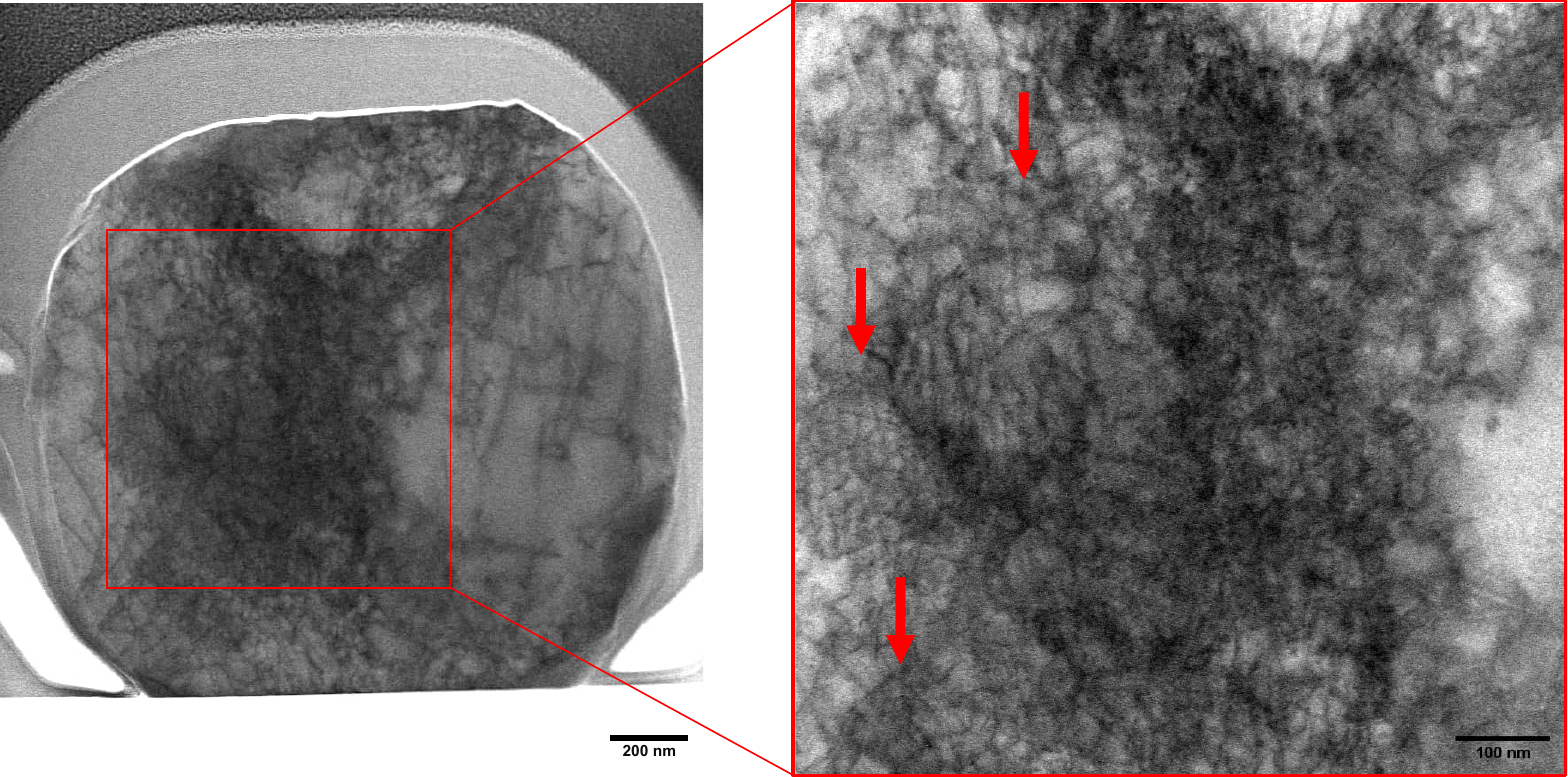}
  \caption{STEM images of post-deformed particles at 1000 $\mathrm{s^{-1}}$ Red arrows identify some of the dislocation cells that formed in the particle during deformation at high strain rates.}
  \label{fig:S9}
\end{figure}

\begin{figure}[h!]
\centering
  \includegraphics[width=0.9\linewidth]{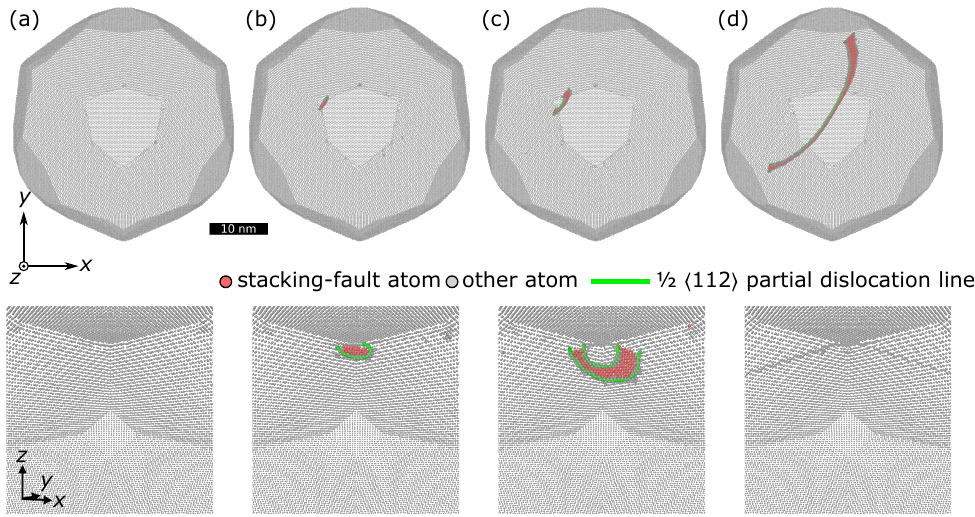}
  \caption{Snapshots of the dislocation nucleation in the nanoparticle with 40 nm diameter. All fcc atoms were deleted, leaving only surface atoms (rendered smaller for better visibility), stacking-fault atoms, and the dislocations. The top row shows a view from the compression direction, the bottom row is a view from inside the particle onto the corner where the first dislocation nucleates. (a) The sample before any dislocation activity. (b) At first, only a leading partial nucleates, but it is pinned at the corner. This situation persists until the stress has further increased. (c)–(d) The leading partial detaches and a trailing partial follows quickly. This is identified as the yield point.}
  \label{fig:S10}
\end{figure}

\begin{figure}[h!]
\centering
  \includegraphics[width=0.6\linewidth]{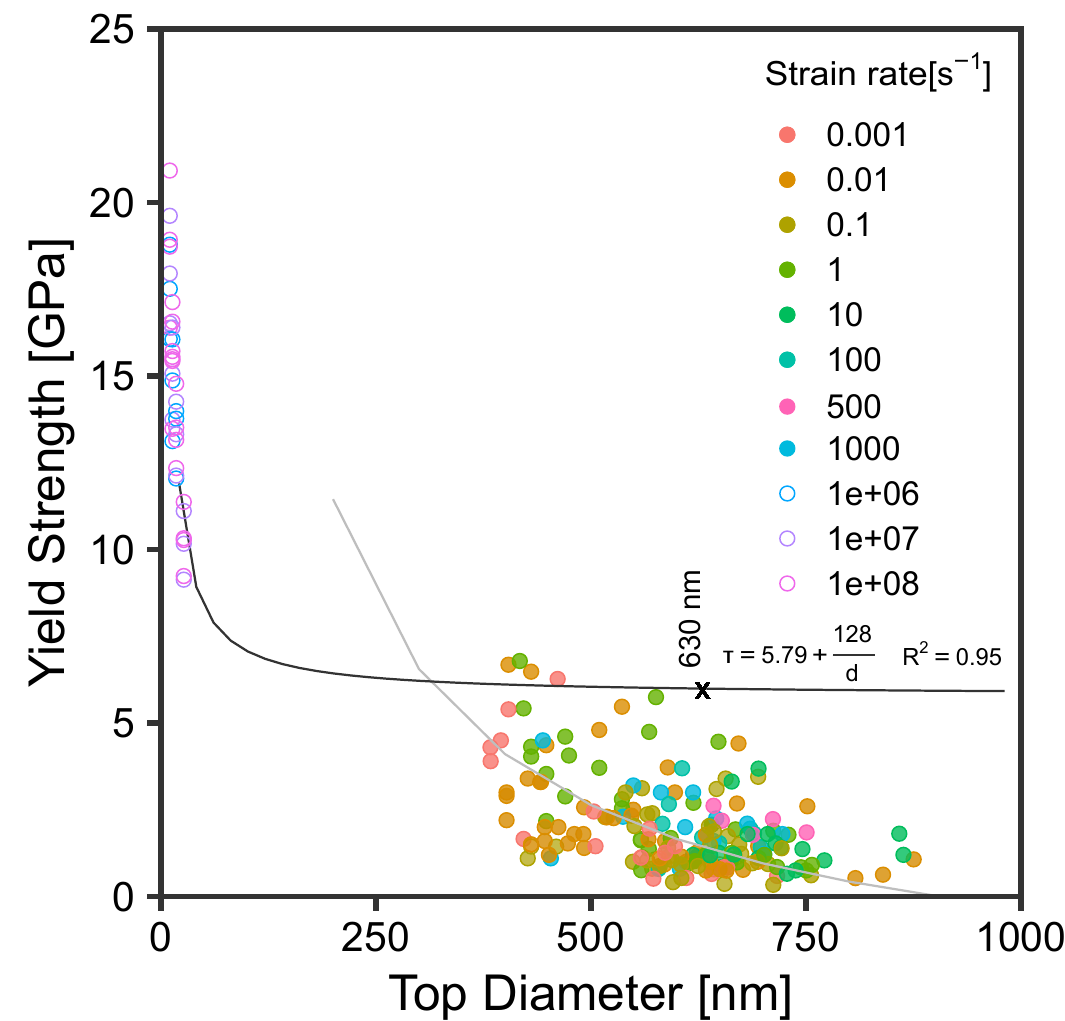}
  \caption{Size effect plot: Yield Strength in GPa vs. Top diameter in nm of particles in experiments (closed symbols) and simulations (open symbols). The fitting line corresponds with Equation (3) in the main text. }
  \label{fig:S11}
\end{figure}

\begin{figure}[h!]
\centering
  \includegraphics[width=0.7\linewidth]{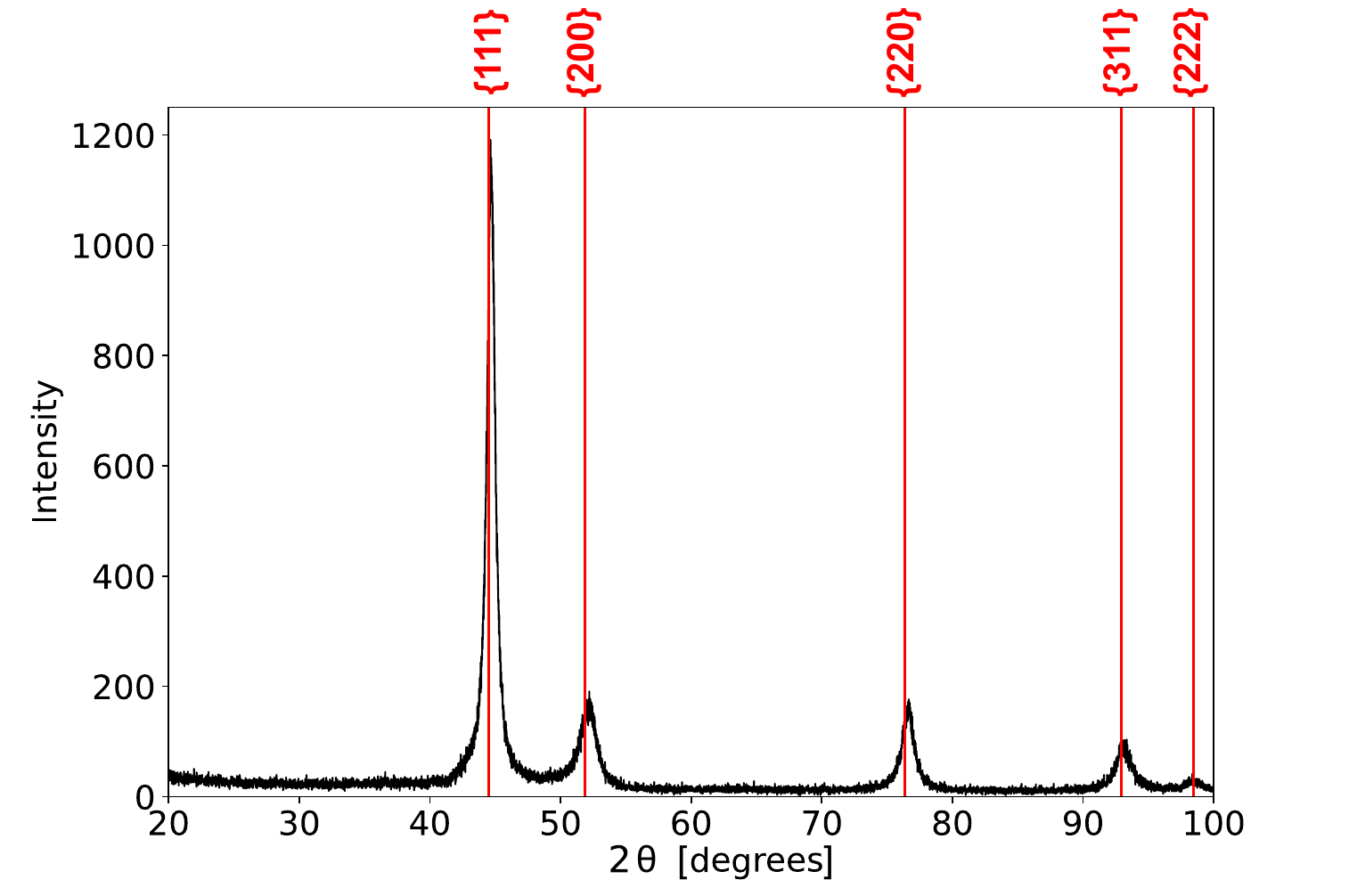}
  \caption{XRD pattern of the deposited nickel film onto the sapphire substrate. The diffraction data shows peaks corresponding to a nickel-textured preferentially in the $\{111\}$ orientation polycrystalline film. The small shifts towards the right could be given by the residual stresses on the film and foreign species that modify the d spacing of the nickel lattice. Reference data used for nickel JCPDS No. 04-0850.}
  \label{fig:S12}
\end{figure}

\begin{figure}[h!]
\centering
  \includegraphics[width=0.7\linewidth]{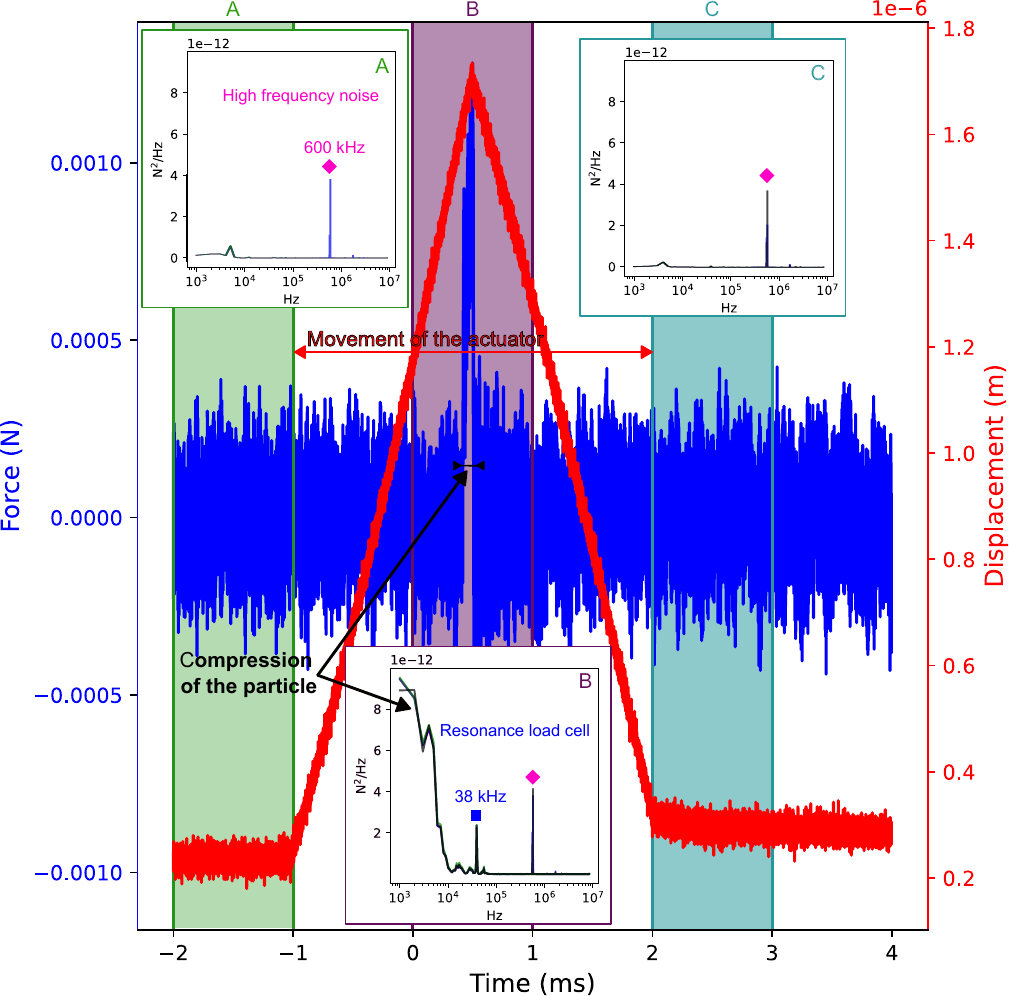}
  \caption{Load (blue) and displacement (red) versus time plot. The insets correspond to the power spectrum density calculated in each of the corresponding highlighted sections. Section A is the one at the beginning of the test, the indenter is recording but the motion has not yet started, this section gives information about the resonant frequencies due to the electronics as nothing is moving. Section B corresponds to the test and the time before and after the test, this section gives information about the frequencies excited during and after the test. Section C is at the end of the test, when there is no more movement, only the same high frequency noise as in the first section can be observed here, which means that the resonant frequencies excited by the movement of the actuator have died down. It should be noted that only the high frequency noise was filtered out with a low pass filter.}
  \label{fig:S13}
\end{figure}

\begin{figure}[h!]
\centering
\includegraphics[width=0.9\linewidth]{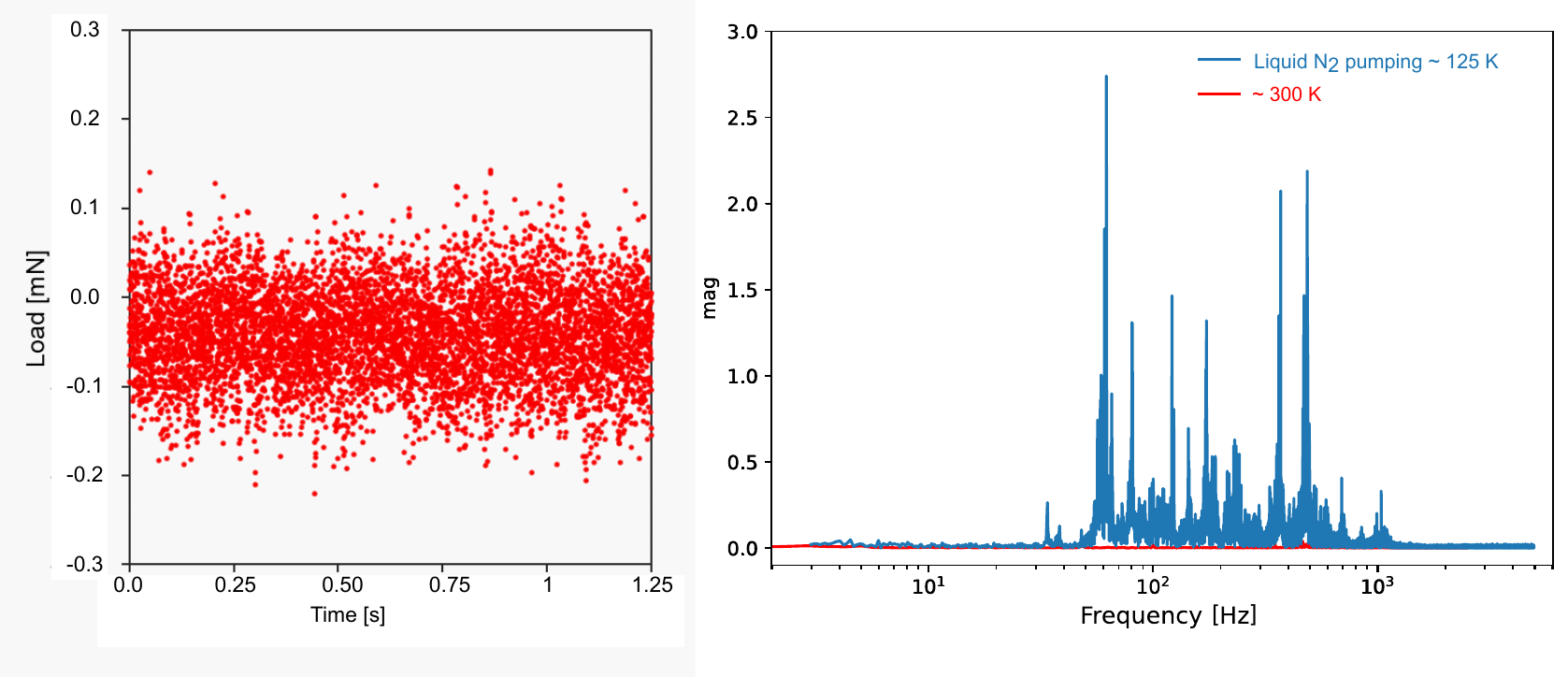}
\caption{Left: Zero load response of the indenter in static conditions at room temperature. Right: Corresponding frequency response (fast Fourier transform (FFT)) of the load at room temperature (red) from image on the left and equivalent at cryogenic temperatures (blue) with the liquid N$\mathrm{_2}$ pump running. The frequencies of these peaks have been characterized using the signal processing library within SciPy and utilized to filter the data at cryogenic temperatures}
\label{fig:S14}
\end{figure}

\begin{figure}[h!]
\centering
  \includegraphics[width=\linewidth]{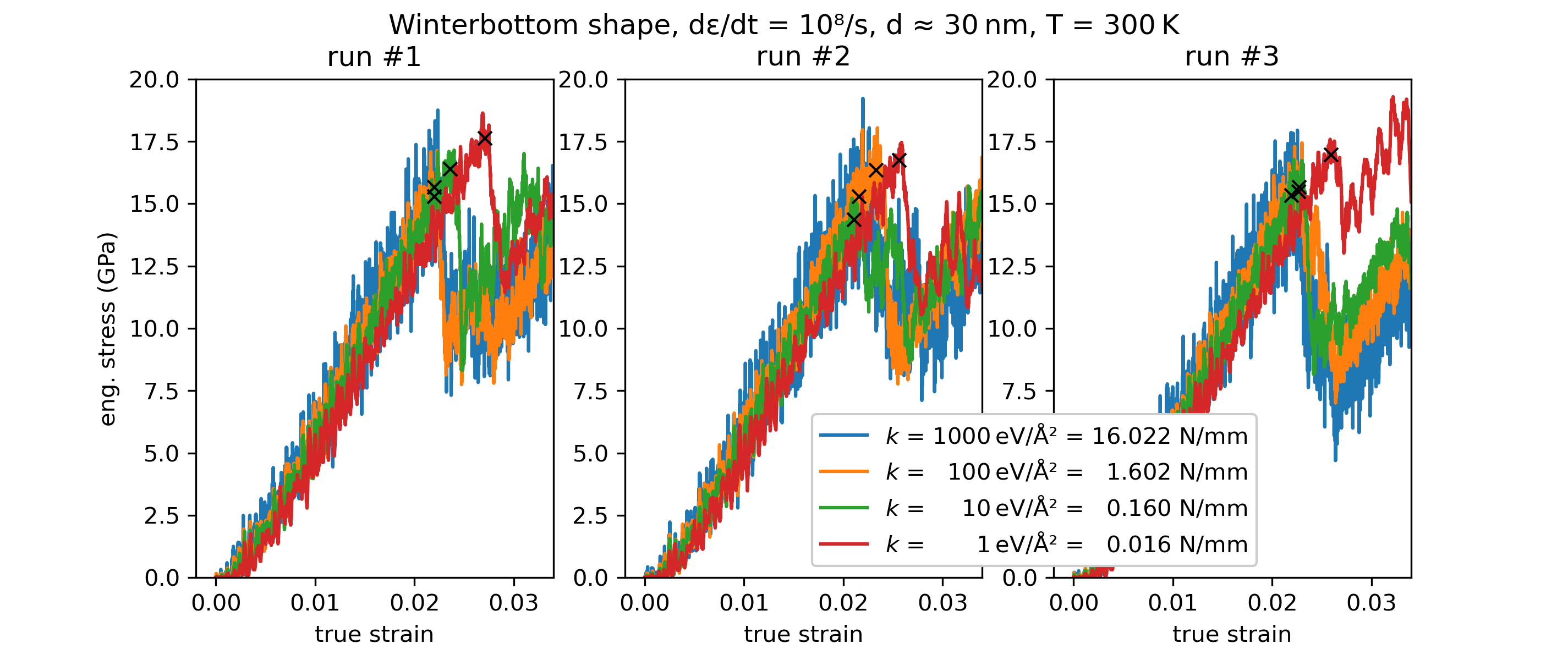}
  \caption{A technical parameter is the stiffness of the harmonic wall used to compress the particles. We tested several spring parameters k and found that this local contact stiffness has negligible influence on the results. Only for a very soft wall with k = 1 eV $\mathrm{\AA^{-2}}$ does the change exceed the fluctuations.}
  \label{fig:S15}
\end{figure}

\begin{figure}[h!]
\centering
  \includegraphics[width=\linewidth]{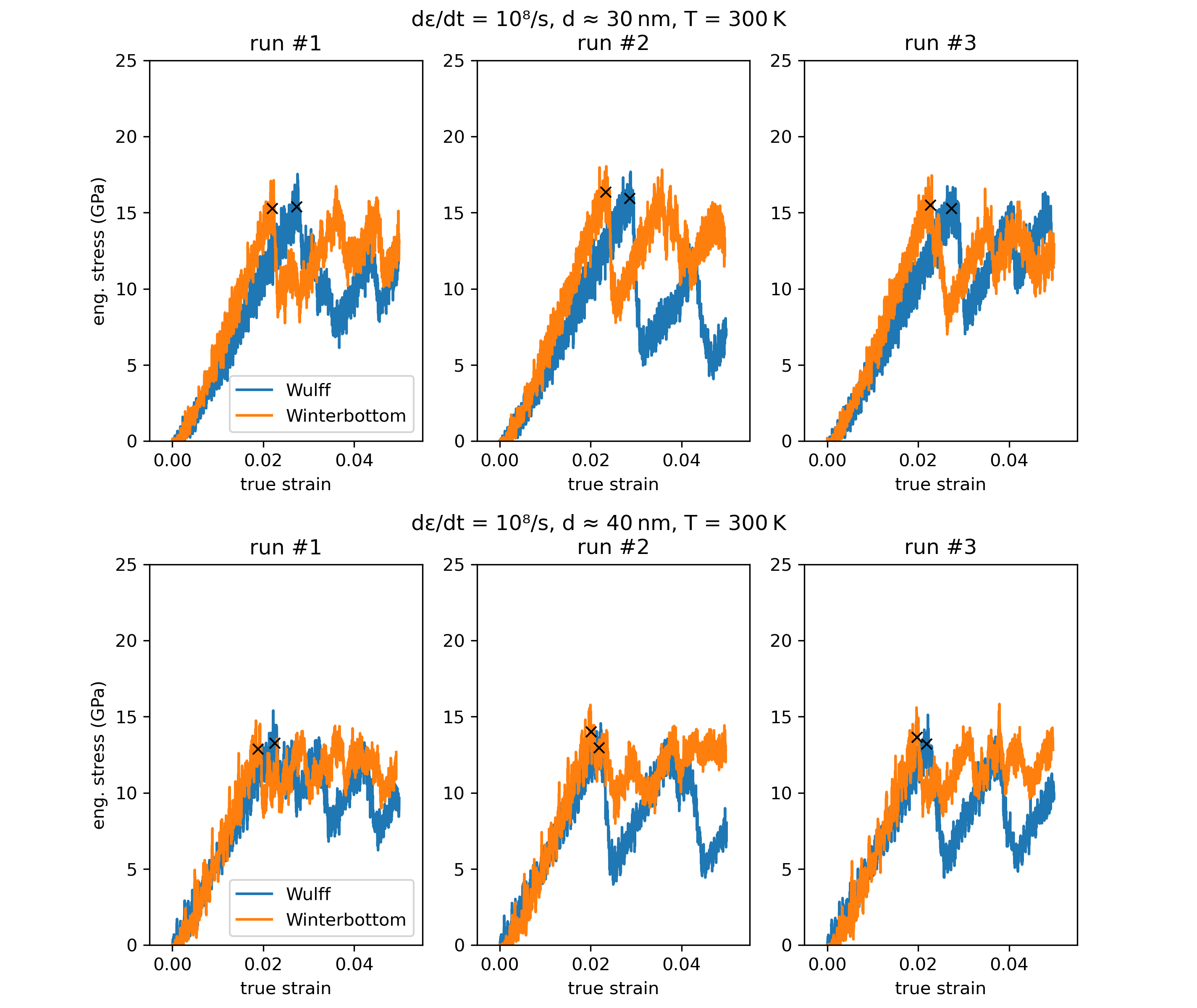}
  \caption{Comparison of simulations with different shapes. The blue curves represent the normal Wulff shape, while the Winterbottom shape is a Wulff shape cut off below a (111) plane to match the experimental shape. The yield stress is the same (within the fluctuations), although the apparent stiffness is slightly affected, especially for the smaller particles. This is due to the nonuniform stress distribution in the particles.}
  \label{fig:S16}
\end{figure}

\begin{figure}[h!]
\centering
  \includegraphics[width=\linewidth]{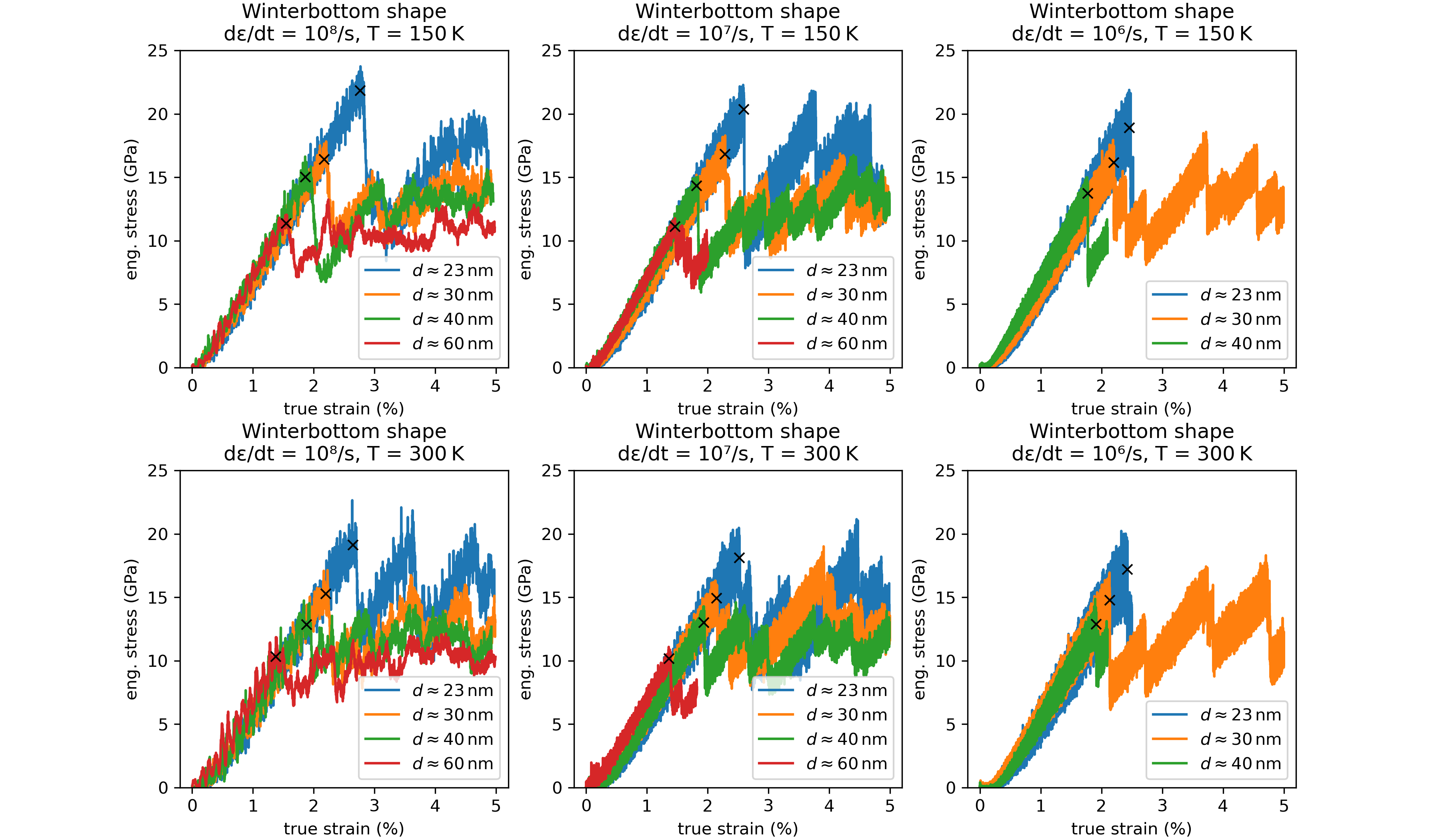}
  \caption{Stress-strain curves for the simulations. Only one of the three statistically independent runs is shown (the other runs yield equivalent results). The crosses are the yield stresses as detected by our algorithm (see Methods).}
  \label{fig:S17}
\end{figure}

\end{document}